\documentclass[prd,aps,nofootinbib,showpacs,showkeys]{revtex4}
\usepackage{graphicx}
\usepackage{bm}
\usepackage{amssymb}
\usepackage{amsmath}
\usepackage{euscript}
\makeatletter
\renewcommand{\@cite}[2]{{#1\if@tempswa , #2\fi}}
\makeatother

\begin{document}

\title{Gauge dependence of effective gravitational field.II.\\
Point-like measuring device}

\author{Taras~S.~Gribouk}\email{taras@theor.phys.msu.ru}
\author{Kirill~A.~Kazakov}\email{kirill@theor.phys.msu.ru}
\author{Petr~I.~Pronin}\email{petr@theor.phys.msu.ru}
\affiliation{Department of Theoretical Physics, Physics Faculty,
Moscow State University, $119899$, Moscow, Russian Federation}

\begin{abstract}
The role of the measurement process in resolving the gauge
ambiguity of the effective gravitational potential is reexamined.
The motion of a classical point-like particle in the field of an
arbitrary linear source, and in the field of another point-like
particle is investigated. It is shown that in both cases the value
of the gravitational field read off from the one-loop effective
action of the testing particle depends on the Feynman weighting
parameter. The found dependence is essential in that it persists
in the expression for the gravitational potential. This result
disproves the general conjecture about gauge independence of the
effective equations of motion of classical point-like particles.
\end{abstract}
\pacs{04.60.Ds,11.10.Lm} \keywords{Gauge dependence, radiative
corrections, measurement, Slavnov identities}

\maketitle

\section{introduction}

The so-called gauge dependence problem is probably the main
obstacle for a straightforward physical interpretation of
calculations in the theory of effective fields. Formulated in
terms of the quantum fields averages, this theory provides general
means for analyzing radiative corrections to various physical
processes including those outside the scope of the standard
scattering theory. In short, the problem consists in the
following. The results of calculations in the effective field
formalism are generally ambiguous because of an arbitrariness in
the choice of gauge conditions fixing the gauge freedom (closely
related to this is the parameterization dependence problem
originating from the freedom in the choice of dynamical variables
in terms of which the theory is quantized). Formal features of the
gauge dependence problem in the case of gravity are quite similar
to those in the Yang-Mills theories. From the point of view of the
Batalin-Vilkovisky formalism \cite{batvil}, for instance, a change
of the gauge conditions induces an anti-canonical transformation
of the effective action in both cases \cite{tyutin}. From the
physical point of view, however, interpretations of this problem
are quite different. Unlike the case of Yang-Mills theory, the
notion of a gauge in the theory of gravitation is directly related
to the properties of physical spacetime. Namely, arbitrariness in
the choice of gauge conditions corresponds to the arbitrariness in
the choice of the reference frame, {\it i.e.,} of the way various
spacetime points are identified by means of the reference bodies.
Accordingly, the gauge dependence problem in quantum gravity is
actually the question of whether a change of the gauge can be
interpreted as a deformation of the reference frame, and {\it vice
versa}.

The notion of the reference frame is closely related to the notion
of measurement, and as such it has an essentially classical
meaning. Therefore, the freedom in the choice of reference frame
in quantum gravity is the same as in classical theory. Thus, the
above question is whether quantization of the gravitational field
introduces an ambiguity into the correspondence between gauge
conditions and reference frames. This question is rather
nontrivial because of the following circumstance. In quantum
theory, the gauge fixing procedure, {\it e.g.,} in the functional
integral approach, is formulated in terms of the integration
variables, rather than expectation values used in the discussion
of physical issues such as transitions between different reference
frames. Detailed analysis shows that this rather indirect
connection between gauge conditions and reference frames is indeed
unambiguous as far as contributions of zero order in the Planck
constant $\hbar$ are considered \cite{kazakov1,kazakov2}. More
precisely, this connection turns out to be the same as in
classical theory up to a spacetime diffeomorphism.

At the next order in $\hbar,$ things turn out to be more
complicated. As was emphasized in Ref.~\cite{dalvit1}, a
consistent treatment of the gauge dependence of $O(\hbar)$-
corrections to the effective gravitational field requires an
explicit introduction of a measuring device into consideration. In
the process of measurement, the value of the gravitational field
is affected by the measuring device. Even if the mass of the
measuring apparatus is infinitely small, and therefore, so is its
contribution to the gravitational field, it cannot be neglected
nevertheless. The point is that the value of the effective
gravitational field is read off from the effective action of the
measuring apparatus (or from its effective equations of motion),
which is also small, while the relative value of the
$O(\hbar)$-corrections to the gravitational field is mass-
independent. Thus, the $O(\hbar)$ contribution of the
device--graviton interaction to the effective device action turns
out to be of the same order as that of the graviton interaction
with the external source of the gravitational field. It was shown
in Ref.~\cite{dalvit1} by an explicit calculation that in the
course of derivation of the effective equations of motion of the
measuring device, the gauge-dependent parts of the two
contributions cancel each other.

In connection with this result, we would like to note the
following. The gauge dependence cancellation was demonstrated in
Ref.~\cite{dalvit1} in a very special case when the source of the
gravitational field as well as the measuring device are point-like
non-relativistic particles. To the best of our knowledge, it has
never been verified under more general conditions. Nevertheless,
the statement about eventual gauge independence of the effective
equations of motion of a point-like measuring device has been used
in a number of later developments as a well established result. In
particular, it has been assumed in investigation of the graviton
corrections to the Maxwell equations \cite{dalvit2}, as well as
particle dynamics in the Robertson Walker spacetime
\cite{dalvit3}. At the same time, it was shown in
Ref.~\cite{kazakov3} that in the case when the role of the
measuring device is played by a classical scalar field, its
effective equations of motion are still gauge dependent. This
raises a question about factual conditions under which the gauge
dependence cancellation takes places. The purpose of this paper is
to reexamine the role of the measurement using point-like
classical particles in solving the gauge dependence problem. The
reason we turn back to this case is that although the class of
gauge conditions considered in Ref.~\cite{dalvit1} is general
enough to assert that the found cancellation of gauge dependence
is not accidental, the whole consideration of Ref.~\cite{dalvit1}
was carried out in the framework of the background field method.
As is well known, the gauge-fixed quantum action in this case
possesses an additional symmetry connected with the gauge
transformations of the background field. It was mentioned already
in Ref.~\cite{kazakov3} that this symmetry reduces the number of
diagrams contributing to the gauge dependent part of the effective
device action. The question of whether the gauge dependence
cancellation is a byproduct of the background field method prompts
one to go beyond this method. From the formal point of view, this
means nothing but a discrete change of the gauge conditions from
the background to the ordinary ones. If the effective device
action is really gauge independent, it must be invariant under
this change. Below, we will follow the general method of
calculating the gauge dependent part of the effective device
action, developed in Ref.~\cite{kazakov3} (thereafter referred to
as I). This method has a number of advantages. First of all, it
makes an explicit evaluation of the mean gravitational field
unnecessary. Second, in the case when the device contribution to
the mean gravitational field is small, which is of primary
interest, it allows one to avoid the necessity of performing the
Legendre transformation from the field sources to the mean fields,
required in constructing the effective device action. Finally, the
use of the Slavnov identities makes the structure of various
contributions to the gauge dependent part of the effective device
action most transparent and easily dealt with. In connection with
the latter we would like to emphasize that the reasons in favor of
using the Slavnov identities lie actually far beyond the matter of
convenience. If the gauge dependence cancellation in the effective
equations of motion is a fundamental property of the classical
device indeed, there must be a formal mathematical reason
underlying this cancellation. Such reason is naturally expected to
have its roots in the original gauge invariance of the classical
theory, expressed in the quantum domain by the Slavnov identities.
As we will see, the use of the Slavnov identities reveals an
unfortunate fact that the gauge dependent contributions of the two
types discussed above {\it tend to add up,} at least partially,
{\it rather than subtract.}

Our paper is organized as follows. In Sec.~\ref{linsource}, we
consider the motion of a point-like apparatus in the gravitational
field produced by an arbitrary source linear in the field. We find
no gauge cancellation in this case. To show that the non-linearity
of realistic sources does not change this conclusion, the special
case of a point-like classical source is investigated in
Sec.~\ref{pointsource}. For this purpose, we generalize the
original method developed in I to the nonlinear case by
introducing an auxiliary source for the BRST-variation of the
matter action, and then find the contribution of the nonlinear
graviton-source interaction to the gauge dependent part of the
effective device action. The results of the work are discussed in
Sec.~\ref{conclud}.

As a rule, we employ condensed notation throughout this paper, and
omit the signs of spacetime integrals, implying that the
integration is done along with summation over repeated indices;
integrals along the world lines of the point particles are written
out explicitly. Also, left derivatives with respect to odd
variables are used. The dimensional regularization of all
divergent quantities is assumed.

\section{Effective gravitational field of a linear source}\label{linsource}

In this section, we consider the motion of a point particle in the
gravitational field produced by a matter source linear in the
graviton field operator, {\it i.e.,} $T^{\mu\nu}h_{\mu\nu},$ where
$h_{\mu\nu} = g_{\mu\nu} - \eta_{\mu\nu}$ is the deviation of the
spacetime metric $g_{\mu\nu}$ from the flat Minkowski metric
$\eta_{\mu\nu},$ and the source $T^{\mu\nu}$ a function of the
spacetime coordinates, satisfying
\begin{eqnarray}\label{conserv}
\frac{\partial T^{\mu\nu}}{\partial x^{\mu}} = 0\,,
\end{eqnarray}
and arbitrary otherwise. We write the ordinary derivatives in
Eq.~(\ref{conserv}) instead of covariant ones implying that all
calculations throughout this paper are carried out in the linear
approximation with respect to $T^{\mu\nu}$ (note that the external
gravitational field is $O(T).$) The mass $m$ of the particle is
assumed sufficiently small so as to neglect its contribution to
the mean gravitational field $\langle h_{\mu\nu}\rangle.$ The
value of the effective gravitational field measured by the
particle is read off from its effective equations of motion
$$\frac{\delta \Gamma_{\rm m}[x(\tau), h(\tau)]}{\delta x^{\mu}(\tau)} = 0\,,$$
where $\Gamma_{\rm m}$ denotes the effective action of the
particle, which is a function of the particles' spacetime position
$x(\tau) \equiv \{x^{\mu}(\tau)\}$ parameterized by its proper
time $\tau$ built from the flat metric, $d\tau^2 = \eta_{\mu\nu}
dx^{\mu} dx^{\nu},$ and of the mean gravitational field $h(\tau)
\equiv \{h_{\mu\nu}[x(\tau)]\}$ at the point $x(\tau);$ the angle
brackets denoting the operation of averaging are omitted for
brevity. The question of gauge (in)dependence of the measured
gravitational field is thus the question of gauge (in)dependence
of its effective action $\Gamma_{\rm m}$.

The gauge dependent part of $\Gamma_{\rm m}$ is the sum of two
different contributions. One of them comes from an explicit gauge
dependence of the effective action $\Gamma[x,h]$, while the other
-- from an implicit gauge dependence of the mean field $h = h(T).$
As was shown in I, the full derivative of the effective device
action with respect to a gauge parameter $\zeta$, which takes into
account both contributions, can be written as
\begin{eqnarray}\label{full}
\frac{d\Gamma_{\rm m}}{d\zeta} = \frac{\partial W_{\rm
m}}{\partial\zeta}\,,
\end{eqnarray}
\noindent where $W_{\rm m}$ is the device contribution to the
generating functional of connected Green functions, $W\,.$ The
latter is defined by
\begin{eqnarray}\label{gener}&&
e^{iW} = {\displaystyle\int}dh dC d\bar{C} \exp\{i (\Sigma +
\bar{\beta}^{\alpha}C_{\alpha} + \bar{C}^{\alpha}\beta_{\alpha} +
T^{\mu\nu}h_{\mu\nu})\}\,,
\end{eqnarray}
\noindent where
\begin{eqnarray}&&\label{sigma}
\Sigma = S_{\rm FP} + \int d^4x
~K^{\mu\nu}D_{\mu\nu}^{\alpha}C_{\alpha} - \int d^4
x~\frac{L^{\gamma}}{2}
f^{\alpha\beta}_{~~~\gamma}C_{\alpha}C_{\beta} + J\int d^4 x
\frac{\delta S_{\rm m}}{\delta h_{\mu\nu}} {D}_{\mu\nu}^{\alpha}
C_{\alpha}\,,
\end{eqnarray}
\noindent $S_{\rm FP}$ is the Faddeev-Popov quantum action,
\begin{eqnarray}\label{fp}
S_{\rm FP} = S + S_{\rm m} + S_{\rm gf} + \int d^4
x~\bar{C}^{\beta}F_{\beta}^{,\mu\nu}D_{\mu\nu}^{\alpha}C_{\alpha}\,;
\end{eqnarray}
\noindent $S, S_{\rm m}$ are the action functionals for the
gravitational field and the measuring apparatus,
respectively,\footnote{Our notation is $R_{\mu\nu} \equiv
R^{\alpha}_{\mu\alpha\nu} =
\partial_{\alpha}\Gamma^{\alpha}_{\mu\nu} - \cdot\cdot\cdot,
~R \equiv R_{\mu\nu} g^{\mu\nu}, ~g\equiv {\rm det}~g_{\mu\nu},
~g_{\mu\nu} = {\rm sgn}(+,-,-,-).$ We use units in which $c =
\hbar = 16\pi G = 1,$ $G$ being the Newton gravitational constant.
Indices are raised and lowered with the help of Minkowski metric
$\eta_{\mu\nu}$.}
\begin{eqnarray}&&\label{actionh}
S = - {\displaystyle\int} d^4 x \sqrt{-g}R\,,
\end{eqnarray}
\noindent
\begin{eqnarray}&&\label{actionm}
S_{\rm m} = - m{\displaystyle\int}
\sqrt{g_{\mu\nu}dx^{\mu}dx^{\nu}} = -
m{\displaystyle\int}d\tau\sqrt{1 + h_{\mu\nu}
\frac{dx^\mu}{d\tau}\frac{dx^\nu}{d\tau}}\ ,
\end{eqnarray}
\noindent while $S_{\rm gf}$ is the gauge-fixing term
\begin{eqnarray}\label{gauge}&& S_{\rm gf} = \int d^4x
\frac{F_{\alpha} F^{\alpha}}{2\xi}\,,
\end{eqnarray}
\noindent where $\xi$ is a parameter weighting the gauge
conditions $F_{\alpha}$ specified below; the c-functions
$K^{\mu\nu}(x)$ (odd) and $L^{\alpha}(x)$ (even) are sources for
the Becchi-Rouet-Stora-Tyutin (BRST) transformations of the
gravitational field and the Faddeev-Popov ghost field
$C_{\alpha},$ respectively \cite{zinnjustin}; $D,$ $\tilde{D}$ are
the generators of gauge transformations of the variables $h,$ $x$
\begin{eqnarray}\label{gaugesym} \delta h_{\mu\nu} &=&
\xi^{\alpha}\partial_{\alpha}h_{\mu\nu} + (\eta_{\mu\alpha} +
h_{\mu\alpha})\partial_{\nu}\xi^{\alpha} + (\eta_{\nu\alpha} +
h_{\nu\alpha})\partial_{\mu}\xi^{\alpha} \equiv
D_{\mu\nu}^{\alpha}\xi_{\alpha}\,, \nonumber\\ \delta x^{\mu} &=&
- \xi^{\mu}\equiv \tilde{D}^{\mu}_{\alpha}\xi^{\alpha}\,,
\end{eqnarray}
\noindent where $\xi^{\alpha}$ are infinitesimal gauge functions;
following I, we introduced into $\Sigma$ the constant source $J$
(odd) for the BRST-variation\footnote{To avoid appearance of the
second derivatives of $x^{\mu}(\tau)$ in the effective apparatus
action, we have written this variation with respect to
$h_{\mu\nu},$ rather than $x^{\mu}$ (as in I). In view of the
gauge invariance of the action $S_{\rm m},$ this amounts simply to
the change $J\to -J\,.$ Accordingly, this change is made in the
Slavnov identity (\ref{slavnov}).} of the apparatus action;
finally, $\bar{\beta}$ and $\beta$ are the ordinary sources for
the Faddeev-Popov ghost--anti-ghost fields, $C$ and $\bar{C},$
respectively.

As was mentioned above, the gauge dependence problem is in fact
the question whether a change of the gauge conditions can be
unambiguously interpreted as a deformation of the reference frame.
Following I, this question will be examined below in the case when
the deformation is induced by a variation of the gauge parameter
$\xi.$ This parameter plays the role of a weight of the set
$F_{\alpha}$ of the gauge conditions, and is a potential source of
gauge ambiguity in the values of observable quantities. In
classical theory, the gravitational field is $\xi$-independent. At
the $\hbar^0$-order of quantum theory, variations of this
parameter induce spacetime diffeomorphisms, thus preserving the
$\xi$-independence of observables \cite{kazakov1,kazakov2}. To see
whether this is so at the first order in $\hbar,$ we have to
calculate, according to Eq.~(\ref{full}), the partial derivative
$\partial W_{\rm m}/\partial\xi.$ This quantity can be
conveniently found using the Slavnov identity for the functional
$W$ modified by adding a source term $Y\int
d^4x~\bar{C}^{\alpha}F_{\alpha},$ $Y$ being a new constant odd
parameter \cite{nielsen}. Denoting the modified generating
functional of connected Green functions by $\mathfrak{W}\,,$ we
thus have
\begin{eqnarray}\label{genernew}&&
e^{i\mathfrak{W}} = {\displaystyle\int}dh dC d\bar{C} \exp\{i
(\Sigma + Y\bar{C}^{\alpha}F_{\alpha} +
\bar{\beta}^{\alpha}C_{\alpha} + \bar{C}^{\alpha}\beta_{\alpha} +
T^{\mu\nu}h_{\mu\nu})\}\,.
\end{eqnarray}
\noindent Except for the last term, the functional $\Sigma$ is
invariant under the following BRST transformation \cite{brst}
\begin{eqnarray}\label{brsta}
\delta h_{\mu\nu} &=&
D_{\mu\nu}^{\alpha}C_{\alpha}\lambda\,, \\
\delta C_{\gamma} &=& -
\frac{1}{2}f^{\alpha\beta}_{~~~\gamma}C_{\alpha}C_{\beta}\lambda\,,
\\ \delta \bar{C}^{\alpha} &=&
\frac{1}{\xi}F^{\alpha}\lambda\,, \label{brstbarc}\\
\label{brstb} \delta x^{\mu} &=&
\tilde{D}^{\mu}_{\alpha}C^{\alpha}\lambda\,,
\end{eqnarray}
where $\lambda$ is a constant (odd) parameter. On the other hand,
the $J$-term is invariant under the ``quantum''
BRST-transformation described by
Eqs.~(\ref{brsta})--(\ref{brstbarc}). Taking into account these
facts, it is not difficult to derive the following Slavnov
identity for the functional $\mathfrak{W}$ (see Appendix of I)
\begin{eqnarray}\label{slavnov}&&
T^{\mu\nu}\frac{\delta \mathfrak{W}}{\delta K^{\mu\nu}} -
\bar{\beta}^{\alpha}\frac{\delta \mathfrak{W}}{\delta L^{\alpha}}
- \frac{1}{\xi} \beta_{\alpha}F^{\alpha,\mu\nu}\frac{\delta
\mathfrak{W}}{\delta T^{\mu\nu}} + \frac{\partial
\mathfrak{W}}{\partial J} - Y \beta_{\alpha}\frac{\delta
\mathfrak{W}}{\delta\beta_{\alpha}} - 2 Y\xi\frac{\partial
\mathfrak{W}}{\partial\xi} = 0\,.
\end{eqnarray}
\noindent Setting $L = \beta = \bar{\beta} = 0$ in
Eq.~(\ref{slavnov}), and collecting terms proportional to $Y$
yields
\begin{eqnarray}\label{slavnov1}
2 \xi\frac{\partial W}{\partial\xi} = - T^{\mu\nu}\frac{\delta
\EuScript{W}}{\delta K^{\mu\nu}} -
\frac{\partial\EuScript{W}}{\partial J}\,,
\end{eqnarray}
\noindent where $W, \EuScript{W}$ are defined by $$\mathfrak{W} =
W + Y \EuScript{W},$$ and the sources $K^{\mu\nu}, J$ are also set
equal to zero after differentiation. Finally, extracting the
device contribution in Eq.~(\ref{slavnov1}), we obtain the Slavnov
identity we are looking for
\begin{eqnarray}\label{slavnov2}
2 \xi\frac{\partial W_{\rm m}}{\partial\xi} = -
T^{\mu\nu}\frac{\delta \EuScript{W}_{\rm m}}{\delta K^{\mu\nu}} -
\frac{\partial\EuScript{W}_{\rm m}}{\partial J}\,.
\end{eqnarray}

Let us proceed to evaluation of the right hand side of
Eq.~(\ref{slavnov2}. In the linear approximation with respect to
the external field $T^{\mu\nu},$ the one-loop contributions of the
first and second term are given by the diagrams pictured in
Figs.~\ref{fig1} and \ref{fig2}, respectively. Some of these
diagrams are zeros identically. Indeed, those pictured
collectively in Fig.~\ref{fig1}(f) vanish in view of the
conservation law, Eq.~(\ref{conserv}). Furthermore, the loop in
the diagrams \ref{fig1}(e) and \ref{fig2}(g) has zero external
momentum flow. Since this loop involves only massless propagators,
its dimensionally regularized value is zero [in the diagram of
Fig.~\ref{fig1}(e), there is also the ghost propagator at zero
momentum attached to the loop, leading to a $0/0$-type
indefiniteness. However, it can be easily resolved to give zero
(see I for more detail)].

Calculation of the remaining diagrams can be simplified using the
ordinary Slavnov identities as follows. Neglecting device
contribution in Eq.~(\ref{slavnov}), differentiating it twice with
respect to $\beta_{\alpha}$ and $T^{\mu\nu}$, and setting all the
sources equal to zero yields
\begin{eqnarray}\label{slavnov3}
\frac{\delta^2 W}{\delta\beta_{\alpha}\delta K^{\mu\nu}} -
\frac{1}{\xi} F^{\alpha,\sigma\lambda} \frac{\delta^2 W}{\delta
T^{\sigma\lambda}\delta T^{\mu\nu}} = 0.
\end{eqnarray}
At the tree level, it reads
\begin{eqnarray}\label{slavnov4}&&
\frac{1}{\xi}F^{\alpha,\mu\nu} G_{\mu\nu\sigma\lambda} =
D^{(0)\beta}_{\sigma\lambda}\tilde{G}_{\beta}^{\alpha}\,,
\\&& ~~D^{(0)\alpha}_{\mu\nu} \equiv
D^{\alpha}_{\mu\nu}(0)\,, \nonumber
\end{eqnarray}
and is easily verified with the help of explicit expressions for
the graviton and ghost propagators, $G_{\mu\nu\sigma\lambda},$
$\tilde{G}_{\alpha}^{\beta}$, given below [see Eqs.~(\ref{g1}),
(\ref{gt1})]. This identity allows us to substitute the graviton
propagator going out of the $Y$-vertex by the ghost propagator.
Furthermore, after having done this substitution, the triple
graviton vertex appearing in the diagrams \ref{fig1}(c),
\ref{fig2}(a,f) can be conveniently expressed through the second
variation of the gravity action with the help of the identity
\begin{eqnarray}&&\label{triple}
\left.\frac{\delta^3 S}{\delta h_{\mu\nu}\delta
h_{\sigma\lambda}\delta h_{\rho\tau}}\right|_{h = 0}
D^{(0)\alpha}_{\mu\nu} + \left.\frac{\delta^2 S}{\delta
h_{\mu\nu}\delta h_{\rho\tau}}\right|_{h = 0} \frac{\delta
D^{\alpha}_{\mu\nu}}{\delta h_{\sigma\lambda}} +
\left.\frac{\delta^2 S}{\delta h_{\mu\nu}\delta
h_{\sigma\lambda}}\right|_{h = 0} \frac{\delta
D^{\alpha}_{\mu\nu}}{\delta h_{\rho\tau}} = 0\,,
\end{eqnarray}
obtained by double differentiating the basic identity
\begin{eqnarray}&&
\frac{\delta S}{\delta h_{\mu\nu}} D^{\alpha}_{\mu\nu} = 0\,.
\end{eqnarray}
\noindent Next, as in I, it can be shown using the one-loop
identity (\ref{slavnov3}) that the sum of diagrams
\ref{fig2}(c),(d) is equal to the diagram \ref{fig1}(d), so that
it is sufficient to find the contribution of the latter. Finally,
applying identity (\ref{slavnov4}) to the graviton line going out
from the $Y$-vertex in the diagrams of Fig.~\ref{fig2}(h), one
sees that diagrams of this type do not contribute because of the
energy-momentum conservation (\ref{conserv}).

The building blocks of the diagrams in the momentum space are
defined as follows. The second variation of the Einstein action
\begin{eqnarray}&&
S^{,\mu\nu~\sigma\lambda}(k) =
\left\{\frac{1}{4}(\eta^{\mu\sigma}\eta^{\nu\lambda} +
\eta^{\mu\lambda}\eta^{\nu\sigma} - 2
\eta^{\mu\nu}\eta^{\sigma\lambda})k^2 + \frac{1}{2}
\left(\eta^{\sigma\lambda}k^\mu k^\nu + \eta^{\mu\nu}k^\sigma
k^\lambda\right) \right. \nonumber\\&& \left. -
\frac{1}{4}\left(\eta^{\sigma\mu} k^\lambda k^\nu +
\eta^{\lambda\mu} k^\sigma k^\nu + \eta^{\sigma\nu} k^\lambda
k^\mu + \eta^{\lambda\nu} k^\sigma k^\mu \right)\right\}\,;
\nonumber
\end{eqnarray}
\noindent Below, we choose the gauge conditions to be
$$F_{\alpha} = \partial^{\mu} h_{\mu\alpha} -
\frac{1}{2}\partial_{\alpha} h\,, \qquad h \equiv \eta^{\mu\nu}
h_{\mu\nu}\,.$$ Then the graviton propagator,
$G_{\mu\nu\alpha\beta},$ and the ghost propagator,
$\tilde{G}^{\alpha}_{\beta},$ defined by
\begin{eqnarray}
(S + S_{\rm gf})^{,\mu\nu~\sigma\lambda}
G_{\sigma\lambda\alpha\beta} &=& -
\delta_{\alpha\beta}^{\mu\nu}\,,\qquad
\delta_{\alpha\beta}^{\mu\nu} =
\frac{1}{2}(\delta_{\alpha}^{\mu}\delta_{\beta}^{\nu} +
\delta_{\alpha}^{\nu}\delta_{\beta}^{\mu})\,, \nonumber\\
F_{\alpha}^{,\mu\nu}D^{(0)\beta}_{\mu\nu}\tilde{G}^{\gamma}_{\beta}
&=& - \delta_{\alpha}^{\gamma}\,, \nonumber
\end{eqnarray}
\noindent take the form
\begin{eqnarray}\label{g1}
G_{\mu\nu\sigma\lambda}(k) = &-&
(\eta_{\mu\sigma}\eta_{\nu\lambda} +
\eta_{\mu\lambda}\eta_{\nu\sigma} - \eta_{\mu\nu}
\eta_{\sigma\lambda}) \frac{1}{k^2} \nonumber\\
&-&(\xi - 1)(\eta_{\mu\sigma} k_{\nu} k_{\lambda} +
\eta_{\mu\lambda} k_{\nu} k_{\sigma} + \eta_{\nu\sigma} k_{\mu}
k_{\lambda} + \eta_{\nu\lambda} k_{\mu}
k_{\sigma})\frac{1}{k^4}\,,
\end{eqnarray}
\noindent and
\begin{eqnarray}&&\label{gt1}
\tilde{G}^{\alpha}_{\beta} =
\frac{\delta^{\alpha}_{\beta}}{k^{2}}\,,
\end{eqnarray}
\noindent respectively. Finally, the graviton-apparatus vertices
are generated by expanding the apparatus action (\ref{actionm}) in
powers of $h_{\mu\nu}$
\begin{eqnarray}\label{actionmexp}
S_{\rm m} &=& - m{\displaystyle\int} d \tau \left[1 +
\frac{1}{2}h_{\mu\nu}\dot x^{\mu}\dot
x^{\nu}-\frac{1}{8}h_{\mu\nu} h_{\alpha\beta}\dot x^{\mu}\dot
x^{\nu}\dot x^{\alpha}\dot x^{\beta} +
\cdot\cdot\cdot\right]\nonumber\\
&\equiv&  - m{\displaystyle\int} d \tau + {\displaystyle\int} d^4
y h_{\mu\nu}(y)T_{\rm
m}^{\mu\nu}(y)+\frac{1}{2}{\displaystyle\int} d^4 y
h_{\mu\nu}(y)h_{\alpha\beta}(y)Q_{\rm
m}^{\mu\nu\alpha\beta}(y)+\cdot\cdot\cdot\,,
\end{eqnarray}
\noindent where
\begin{eqnarray}
T_{\rm m}^{\mu\nu}(y) = - \frac{m}{2}{\displaystyle\int} d \tau
\dot x^{\mu}\dot x^{\nu}\delta^4(y-x(\tau))\,, \nonumber
\end{eqnarray}
\begin{eqnarray}
Q_{\rm m}^{\mu\nu\alpha\beta}(y) = \frac{m}{4}{\displaystyle\int}
d \tau \dot x^{\mu}\dot x^{\nu}\dot x^{\alpha}\dot
x^{\beta}\delta^4(y-x(\tau))\,, \qquad \dot{x}^{\mu} \equiv
\frac{d x^{\mu}(\tau)}{d\tau}\,. \nonumber
\end{eqnarray}
\noindent To avoid appearance of the second derivatives of
$x^{\mu}(\tau)$ in the effective device action, an integration by
parts is to be performed in the diagrams \ref{fig2}(a)--(f). After
all the above transformations, the analytical expressions of the
diagrams to be calculated take the form:

$$\mathfrak{I}_j = \int \frac{d^4 p}{(2\pi)^4} I_j(p)\,,$$
\begin{eqnarray}\label{1a}
I_{1(a)}(p) &=& i~\mu^{\varepsilon}{\displaystyle\int}
\frac{d^{4-\varepsilon}k}{(2\pi)^4}T^{\mu\nu}(p)
    Q_{\rm m}^{\tau\rho\sigma\lambda}(-p)
    G_{\tau\rho\chi\theta}(k)
    \tilde{G}^{\alpha}_{\beta}(p+k)
    \xi
    \tilde{G}^{\beta\gamma}(p+k)
\nonumber\\&&
    \times
    \left\{
k_{\alpha}\delta_{\mu\nu}^{\chi\theta}-
\delta_{\mu\alpha}^{\chi\theta}(p_{\nu}+k_{\nu})-
\delta_{\nu\alpha}^{\chi\theta}(p_{\mu}+k_{\mu})
    \right\}
    \left\{
\eta_{\sigma\gamma}(p_{\lambda}+k_{\lambda})+\eta_{\lambda\gamma}
(p_{\sigma}+k_{\sigma})\right\}\,, \\
I_{1(b)}(p) &=& -i~\mu^{\varepsilon}{\displaystyle\int}
\frac{d^{4-\varepsilon}k}{(2\pi)^4}T^{\mu\nu}(p)
    T_{\rm m}^{\varphi\psi}(-p)
    G_{\chi\theta\varphi\psi}(p)
    \tilde{G}^{\gamma}_{\delta}(k)
    \xi
    \tilde{G}^{\delta\zeta}(k)
    \tilde{G}^{\alpha}_{\beta}(p+k)
\nonumber\\&&
    \times\frac{1}{2}
    \left\{
(p^{\beta}+k^{\beta})\eta^{\tau\rho}
-(p^{\tau}+k^{\tau})\eta^{\beta\rho}
-(p^{\rho}+k^{\rho})\eta^{\beta\tau}
    \right\}
    \left\{
        \eta_{\sigma\zeta}k_{\lambda}+\eta_{\lambda\zeta}k_{\sigma}
    \right\}
\nonumber\\&&
    \times
    \left\{
-p_{\gamma}\delta_{\tau\rho}^{\chi\theta}
-\delta_{\tau\gamma}^{\chi\theta}k_{\rho}
-\delta_{\rho\gamma}^{\chi\theta}k_{\tau}
    \right\}
    \left\{
k_{\alpha}\delta_{\mu\nu}^{\sigma\lambda}
-\delta_{\mu\alpha}^{\sigma\lambda}(p_{\nu}+k_{\nu})
-\delta_{\nu\alpha}^{\sigma\lambda}(p_{\mu}+k_{\mu})
    \right\}\,,
\label{1b}\\
I_{1(c)}(p) &=&
-i~\mu^{\varepsilon}{\displaystyle\int}\frac{d^{4-\varepsilon}
k}{(2\pi)^4}T^{\mu\nu}(p)
    T_{\rm m}^{\kappa\omega}(-p)
    \tilde{G}^{\alpha}_{\beta}(p+k)
    \xi
    \tilde{G}^{\beta\gamma}(p+k)
\nonumber\\&&
    \times
    \left
        (S^{,\sigma\lambda~\tau\rho}(p)
        \left\{
            -k_{\gamma}
\delta_{\sigma\lambda}^{\chi\theta}+\delta_{\sigma\gamma}^{\chi\theta}
(p_{\lambda}+k_{\lambda})+
\delta_{\lambda\gamma}^{\chi\theta}(p_{\sigma}+k_{\sigma})
        \right\}
    \right.
\nonumber\\&&
    \left.
        +S^{,\sigma\lambda~\chi\theta}(k)
        \left\{
-p_{\gamma}\delta_{\sigma\lambda}^{\tau\rho}
+\delta_{\sigma\gamma}^{\tau\rho} (p_{\lambda}+k_{\lambda})+
\delta_{\lambda\gamma}^{\tau\rho}(p_{\sigma}+k_{\sigma})
        \right\}
    \right)
\nonumber\\&&
    \times
    G_{\chi\theta\varphi\psi}(k)
    G_{\tau\rho\kappa\omega}(p)
    \left\{
k_{\alpha}\delta_{\mu\nu}^{\varphi\psi}-\delta_{\mu\alpha}^{\varphi\psi}
        (p_{\nu}+k_{\nu})-\delta_{\nu\alpha}^{\varphi\psi}(p_{\mu}+k_{\mu})
    \right\}\,,
\label{1c}\\
I_{1(d)}(p) &=&
    -i~\mu^{\varepsilon}{\displaystyle\int}\frac{d^{4-\varepsilon}
k}{(2\pi)^4}T^{\mu\nu}(p)
    T_{\rm m}^{\sigma\lambda}(-p)
    \tilde{G}^{\alpha}_{\beta}(p+k)
    G_{\kappa\omega\chi\theta}(k)
    \tilde{G}^{\gamma}_{\delta}(p)
    \xi
    \tilde{G}^{\delta\zeta}(p)
\nonumber\\&&
    \times
    \left\{
k_{\alpha}\delta_{\mu\nu}^{\chi\theta}-\delta_{\mu\alpha}^{\chi\theta}
        (p_{\nu}+k_{\nu})-\delta_{\nu\alpha}^{\chi\theta}(p_{\mu}+k_{\mu})
    \right\}
    \left\{
-k_{\gamma}\delta_{\tau\rho}^{\kappa\omega}
-\delta_{\tau\gamma}^{\kappa\omega}
        p_{\rho}-\delta_{\rho\gamma}^{\kappa\omega}p_{\tau}
    \right\}
\nonumber\\&&
    \times \frac{1}{2}
    \left\{
(p^{\beta}+k^{\beta})\eta^{\tau\rho}
-(p^{\tau}+k^{\tau})\eta^{\beta\rho}
-(p^{\rho}+k^{\rho})\eta^{\beta\tau}
    \right\}
    \left\{
        \eta_{\sigma\zeta}p_{\lambda}+\eta_{\lambda\zeta}p_{\sigma}
    \right\}\,,
\label{1d}
\end{eqnarray}

\begin{eqnarray}
I_{2(a)}(p) &=&
     -i~\mu^{\varepsilon}
{\displaystyle\int}\frac{d^{4-\varepsilon}k}{(2\pi)^4}T_{\rm
m}^{\xi\eta}(p)p_{\xi} \tilde{G}_{\eta\delta}(p)
     \tilde{G}^{\alpha}_{\beta}(p+k)
     \xi
     \tilde{G}^{\beta\gamma}(p+k)
\nonumber\\&&
    \times
    \left
    (S^{,\sigma\lambda~\tau\rho}(p)
        \left\{
            -k_{\gamma}\delta_{\sigma\lambda}^{\chi\theta}
+\delta_{\sigma\gamma}^{\chi\theta} (p_{\lambda}+k_{\lambda})
+\delta_{\lambda\gamma}^{\chi\theta}(p_{\sigma}+k_{\sigma})
        \right\}
    \right.
\nonumber\\&&
    \left.
        + S^{,\sigma\lambda~\chi\theta}(k)
            \left\{
                -p_{\gamma}
\delta_{\sigma\lambda}^{\tau\rho}+\delta_{\sigma\gamma}^{\tau\rho}
(p_{\lambda}+k_{\lambda})
+\delta_{\lambda\gamma}^{\tau\rho}(p_{\sigma}+k_{\sigma})
            \right\}
    \right)
\nonumber\\&&
    \times
    h_{\tau\rho}(-p)
    G_{\chi\theta\varphi\psi}(k)
        \left\{
p^{\delta}\eta^{\mu\nu}-p^{\mu}\eta^{\delta\nu}-p^{\nu}\eta^{\delta\mu}
    \right\}
\nonumber\\&&
    \times
    \left\{
        k_{\alpha}\delta_{\mu\nu}^{\varphi\psi} -
\delta_{\mu\alpha}^{\varphi\psi}
        (p_{\nu}+k_{\nu})-\delta_{\nu\alpha}^{\varphi\psi}(p_{\mu}+k_{\mu})
    \right\}\,,
\label{2a}\\
I_{2(b)}(p) &=& -i~\mu^{\varepsilon}
{\displaystyle\int}\frac{d^{4-\varepsilon} k}{(2\pi)^4}T_{\rm
m}^{\xi\eta}(p)p_{\xi}
        \tilde{G}^{\alpha}_{\beta}(p+k)
    \tilde{G}^{\gamma}_{\delta}(k)
    \xi
    \tilde{G}^{\delta\zeta}(k)
    \tilde{G}_{\eta\epsilon}(p)
\nonumber\\&&
     \times\frac{1}{2}
     \left\{
        (p^{\beta} + k^{\beta})\eta^{\tau\rho}-(p^{\tau} +
k^{\tau})\eta^{\beta\rho}
        -(p^{\rho}+k^{\rho})\eta^{\beta\tau}
     \right\}
     \left\{
        \eta_{\sigma\zeta}k_{\lambda}+ \eta_{\lambda\zeta}k_{\sigma}
     \right\}
\nonumber\\&&
    \times
    \left\{
        -p_{\gamma}h_{\tau\rho}(-p)-h_{\tau\gamma}(-p)k_{\rho}-
h_{\rho\gamma}(-p) k_{\tau}
    \right\}
    \left\{
p^{\epsilon}\eta^{\mu\nu}-p^{\mu}\eta^{\epsilon\nu}
-p^{\nu}\eta^{\epsilon\mu}
    \right\}
\nonumber\\&&
    \times
    \left\{
k_{\alpha}\delta_{\mu\nu}^{\sigma\lambda}-\delta_{\mu\alpha}^{\sigma\lambda}
        (p_{\nu} + k_{\nu})-\delta_{\nu\alpha}^{\sigma\lambda}(p_{\mu} +
k_{\mu})
    \right\}\,,
\label{2b}\\
I_{2(e)}(p) &=& -i~\mu^{\varepsilon}{\displaystyle\int}
\frac{d^{4-\varepsilon}k}{(2\pi)^4}
    \tilde{G}^{\gamma}_{\delta}(k)
    \xi
    \tilde{G}^{\delta\zeta}(k)
    \tilde{G}^{\alpha}_{\beta}(p+k)
    \left\{
        \eta_{\sigma\zeta}k_{\lambda}+\eta_{\lambda\zeta}k_{\sigma}
    \right\}
\nonumber\\&&
    \times\frac{1}{2}
    \left\{
(p^{\beta}+k^{\beta})\eta^{\tau\rho}
-(p^{\tau}+k^{\tau})\eta^{\beta\rho}
-(p^{\rho}+k^{\rho})\eta^{\beta\tau}
    \right\}
    \nonumber\\&&
    \times
    \left\{
-p_{\gamma} h_{\tau\rho}(-p) - h _{\tau\gamma}(-p) k_{\rho} -
h_{\rho\gamma}(-p) k_{\tau}
    \right\}\nonumber\\&&
    \times \left\{T_{\rm m}^{\mu\nu}(p)\left[
k_{\alpha}\delta_{\mu\nu}^{\sigma\lambda}
-\delta_{\mu\alpha}^{\sigma\lambda}(p_{\nu}+k_{\nu})
-\delta_{\nu\alpha}^{\sigma\lambda}(p_{\mu}+k_{\mu})
    \right] \right.\nonumber\\&&
    \left. - Q_{\rm m}^{\sigma\lambda\mu\nu}(p)
    \left[\eta_{\mu\alpha}(p_{\nu} + k_{\nu})
    + \eta_{\nu\alpha}(p_{\mu} + k_{\mu})\right]\right\}\,,
\label{2e}\\
I_{2(f)}(p) &=& -i~\mu^{\varepsilon}
{\displaystyle\int}\frac{d^{4-\varepsilon}k}{(2\pi)^4}
\tilde{G}^{\alpha}_{\beta}(p+k)
    \xi
    \tilde{G}^{\beta\gamma}(p+k)
    G_{\chi\theta\omega\xi}(k)
    h_{\tau\rho}(-p)
\nonumber\\&&
    \times
    \left(
        S^{,\sigma\lambda~\tau\rho}(p)
        \left\{
            -k_{\gamma}
\delta_{\sigma\lambda}^{\chi\theta}
+\delta_{\sigma\gamma}^{\chi\theta}(p_{\lambda}+k_{\lambda})+
            \delta_{\lambda\gamma}^{\chi\theta}(p_{\sigma}+k_{\sigma})
        \right\}
    \right.
\nonumber\\&&
    \left.
        +S^{,\sigma\lambda~\chi\theta}(k)
        \left\{
-p_{\gamma}\delta_{\sigma\lambda}^{\tau\rho}
+\delta_{\sigma\gamma}^{\tau\rho}(p_{\lambda}+k_{\lambda})+
            \delta_{\lambda\gamma}^{\tau\rho}(p_{\sigma}+k_{\sigma})
        \right\}
    \right)
\nonumber\\&& \times \left\{T_{\rm m}^{\mu\nu}(p)\left[
k_{\alpha}\delta_{\mu\nu}^{\omega\xi}
-\delta_{\mu\alpha}^{\omega\xi}(p_{\nu}+k_{\nu})
-\delta_{\nu\alpha}^{\omega\xi}(p_{\mu}+k_{\mu})
    \right] \right.\nonumber\\&&
    \left. - Q_{\rm m}^{\omega\xi\mu\nu}(p)
    \left[\eta_{\mu\alpha}(p_{\nu} + k_{\nu})
    + \eta_{\nu\alpha}(p_{\mu} + k_{\mu})\right]\right\}\,,
\label{2f}
\end{eqnarray}
\noindent where $h_{\mu\nu}(p)$ stands for the tree value of the
gravitational field produced by $T^{\mu\nu}$ (in the linear
approximation), $$h_{\mu\nu}(p) =
G_{\mu\nu\alpha\beta}(p)T^{\alpha\beta}(p)\,,$$ $\mu$ -- arbitrary
mass scale, and $\varepsilon = 4 - d,$ $d$ being the
dimensionality of spacetime.

As in Refs.~\cite{dalvit1} and I, we restrict ourselves to the
calculation of the leading quantum corrections in the low-energy
limit. To this end, it is sufficient to find the ultraviolet
divergences of the above Feynman integrals. The logarithmic part
of the integrals, representing the leading contribution in the
present case, can then be obtained by substituting
\begin{eqnarray}\label{alg}&&
\frac{1}{\varepsilon} \to -
\frac{1}{2}\ln\left(\frac{-p^2}{\mu^2}\right)\,.
\end{eqnarray}
\noindent To find the ultraviolet divergences, one has to expand
the integrands in powers of the momentum transfer, $p_{\mu},$ and
to collect terms proportional to $k^{-4}.$ Obviously, expansion of
the ghost propagators $\tilde{G}_{\alpha}^{\beta}(p+k)$ is needed
only. Since the degree of divergence of the above integrals is
$\le 3,$ it is sufficient to write
\begin{eqnarray}&&
\tilde{G}^{\alpha}_{\beta}(p + k) =
\frac{\delta^{\alpha}_{\beta}}{k^2} \left(1 - \frac{2 (p k)}{k^2}
- \frac{p^2}{k^2} + \frac{4 (p k)^2}{k^4} + \frac{4 (p k)
p^2}{k^4} - \frac{8 (p k)^3}{k^6} +
O\left(\frac{p^4}{k^4}\right)\right)\,. \nonumber
\end{eqnarray}
\noindent The tensor multiplication as well as integration over
angles in the $k$-space can be easily performed using the tensor
package \cite{reduce} for the REDUCE system. Thus, we find
\begin{eqnarray}\label{form}
I_j^{\rm log}(p) =
\frac{m\xi}{16\pi^2}\ln\left(\frac{-p^2}{\mu^2}\right)\int d\tau
\exp\{ip_{\mu}x^{\mu}(\tau)\} \EuScript{I}_j(\tau,p)\,,
\end{eqnarray}
\begin{eqnarray}
\EuScript{I}_{1(a)}(\tau,p) &=& \frac{1}{24}\left\{(2\xi + 1)T(p)
+ 4(\xi-1)\dot{x}^{\mu}\dot{x}^{\nu}T_{\mu\nu}(p)\right\}\,,
\nonumber\\
\EuScript{I}_{1(b)}(\tau,p) &=& \frac{1}{3}\left\{T(p) - (\xi - 2
)\frac{(\dot{x}^{\mu}p_{\mu})^2}{p^2}T(p) \right\}\,, \nonumber\\
\EuScript{I}_{1(c)}(\tau,p) &=& - \frac{1}{6}\left\{(2\xi - 1)T(p)
+ 10\xi\frac{(\dot{x}^{\mu}p_{\mu})^2}{p^2}T(p) + 2(2\xi + 1)
\dot{x}^{\mu}\dot{x}^{\nu}T_{\mu\nu}(p)\right\}\,, \nonumber\\
\EuScript{I}_{1(d)}(\tau,p) &=& - \frac{(\xi-1)}{3}
\frac{(\dot{x}^{\mu}p_{\mu})^2}{p^2}T(p)\,,\nonumber\\
\EuScript{I}_{2(a)}(\tau,p) &=& - \frac{2(2\xi+1)}{3}
\frac{(\dot{x}^{\mu}p_{\mu})^2}{p^2}T(p)\,,\nonumber\\
\EuScript{I}_{2(b)}(\tau,p) &=& \frac{
(\dot{x}^{\mu}p_{\mu})^2}{2 p^2}T(p)\,,\nonumber\\
\EuScript{I}_{2(e)}(\tau,p) &=& \frac{1}{12}\left\{2 T(p) + 4
\dot{x}^{\mu}\dot{x}^{\nu}T_{\mu\nu}(p) + \frac{
(\dot{x}^{\mu}p_{\mu})^2}{p^2}T(p) + 2\frac{
(\dot{x}^{\alpha}p_{\alpha})^2}{p^2}\dot{x}^{\mu}\dot{x}^{\nu}T_{\mu\nu}(p)
\right\}\,,
\nonumber\\
\EuScript{I}_{2(f)}(\tau,p) &=& - \frac{1}{24}\left\{(6\xi - 1)
T(p) - 14\frac{(\dot{x}^{\mu}p_{\mu})^2}{p^2}T(p) + 4\left[(3\xi +
1) - \frac{ (\dot{x}^{\alpha}p_{\alpha})^2}{p^2}\right]
\dot{x}^{\mu}\dot{x}^{\nu}T_{\mu\nu}(p)\right\}\,, \nonumber
\end{eqnarray}
\noindent where $T\equiv T^{\mu\nu}\eta_{\mu\nu}.$ When bringing
the results of the loop integration to the form (\ref{form}), we
have used the identity $\eta_{\mu\nu}\dot{x}^{\mu}\dot{x}^{\nu} =
1,$ and changed $p_{\mu}\to - p_{\mu}$ in the expressions for
$I_{2(a-f)}.$ Doubling the contribution of the diagram
\ref{fig1}(d) and summing up, we finally obtain the following
expression for the full $\xi$-derivative of the effective device
action
\begin{eqnarray}
\frac{d\Gamma^{\rm loop}_{\rm m}}{d\xi} &=& \frac{m}{32\pi^2}\int
d\tau \int \frac{d^4 p}{(2\pi)^4}\exp\{ip_{\mu}x^{\mu}(\tau)\}
\ln\left(\frac{-p^2}{\mu^2}\right) \left\{\frac{2\xi - 3}{4}T(p)
\right.\nonumber\\&& \left. + \frac{3\xi + 1
}{3}\dot{x}^{\mu}\dot{x}^{\nu}T_{\mu\nu}(p) + \frac{24\xi - 11}{6}
\frac{(\dot{x}^{\mu}p_{\mu})^2}{p^2}T(p) -
\frac{1}{3}\frac{(\dot{x}^{\alpha}p_{\alpha})^2}{p^2}
\dot{x}^{\mu}\dot{x}^{\nu}T_{\mu\nu}(p)\right\}\,.
\end{eqnarray}
\noindent Comparison of the right hand side of this equation with
the linear term in the expansion of the apparatus action,
Eq.~(\ref{actionmexp}), shows that the $\xi$-dependent part of the
effective gravitational field cannot be represented in the form
$D^{(0)\alpha}_{\mu\nu}\Xi_{\alpha}$ with some $\Xi_{\alpha}.$
That this implies an ambiguity in the values of physical
quantities is most clearly seen considering the particular case of
a static source and non-relativistic testing particle. In this
case, one has $T^{\mu\nu}(p) =
2\pi\delta(p^0)\mathfrak{T}^{\mu\nu}(\bm{p}),$
$\dot{x}^{\mu}\approx \delta^{\mu}_0,$
$\dot{x}^{\mu}p_{\mu}\approx 0.$ Using the formula
$$\int\frac{d^3\bm{p}}{(2\pi)^3}\exp(i\bm{pr}) \ln\bm{p}^2
= -\frac{1}{2\pi r^3}\,, \qquad r\equiv |\bm{r}|\,,$$ and
restoring the ordinary units yields
\begin{eqnarray}\label{result1}
\frac{d\Gamma^{\rm loop}_{\rm m}}{d\xi} &=& - \frac{G^2\hbar}{\pi
c^5}\int dt\int d^3\bm{r}
\frac{m\mathfrak{T}_{\mu\nu}(\bm{r})}{|\bm{r} - \bm{x}(\tau)|^3}
\left\{(2\xi - 3)\eta^{\mu\nu} + 4\left(\xi +
\frac{1}{3}\right)\delta^{\mu}_0\delta^{\nu}_0 \right\}\,, \\
\mathfrak{T}^{\mu\nu}(\bm{r}) &=&
\int\frac{d^3\bm{p}}{(2\pi)^3}\mathfrak{T}^{\mu\nu}(\bm{p})\exp(-i\bm{pr})\,.
\nonumber
\end{eqnarray}
\noindent It follows from Eq.~(\ref{result1}) that the value of
the static gravitational potential, $V,$ measured by observing the
motion of a non-relativistic point particle, is $\xi$-dependent:
\begin{eqnarray}
\frac{dV}{d\xi} &=& \frac{G^2\hbar}{\pi c^5}\int d^3\bm{r}
\frac{\mathfrak{T}_{\mu\nu}(\bm{r})}{|\bm{r} - \bm{x}(\tau)|^3}
\left\{(2\xi - 3)\eta^{\mu\nu} + 4\left(\xi +
\frac{1}{3}\right)\delta^{\mu}_0\delta^{\nu}_0 \right\}\,.
\nonumber
\end{eqnarray}
\noindent Thus, as in the case of scalar field, an explicit
introduction of the classical measuring apparatus does not remove
the gauge ambiguity of the effective gravitational potential. One
might think, however, that this is because of the artificial form
of the source for the gravitational field. The point is that this
source is not invariant with respect to spacetime diffeomorphisms.
As we mentioned in the Introduction, the reason for the gauge
dependence cancellation, if any, is expected to have its roots in
the gauge invariance of the initial classical theory. In the next
section, therefore, we turn to the investigation of a realistic
source nonlinear in the gravitational field.

\section{Effective gravitational field of a point-like
particle}\label{pointsource}

In this section, we will consider the special case when the
gravitational field is produced by a point-like classical particle
with mass $M.$ This is the setting of Ref.~\cite{dalvit1}, except
that we use the ordinary gauge conditions instead of the
background ones. The method used in the preceding section does not
apply to this case because of the nonlinearity of the source.
However, it can be readily extended to take into account this
nonlinearity as follows.

Let us introduce, analogously to the source $J,$ a constant source
$\EuScript{J}$ (odd) for the BRST-variation of the source particle
action
$$\EuScript{J}\int d^4 x \frac{\delta S_{\rm M}}
{\delta h_{\mu\nu}}{D}^{\alpha}_{\mu\nu}C^{\alpha}\,,$$ where
$S_{\rm M }$ is the source particle action
\begin{eqnarray}&&\label{actionsource}
S_{\rm m} = - M{\displaystyle\int}
\sqrt{g_{\mu\nu}dz^{\mu}dz^{\nu}} = -
M{\displaystyle\int}d\theta\sqrt{1 + h_{\mu\nu}
\frac{dz^\mu}{d\theta}\frac{dz^\nu}{d\theta}}\ , \qquad d\theta^2
= \eta_{\mu\nu}dz^{\mu}dz^{\nu}\ .
\end{eqnarray}
\noindent Expanding this action in powers of $h_{\mu\nu}$
generates the vertices of the source-graviton interaction:
\begin{eqnarray}
S_{\rm M} &=& - M{\displaystyle\int} d \theta \left[1 +
\frac{1}{2}h_{\mu\nu}\dot z^{\mu}\dot
z^{\nu}-\frac{1}{8}h_{\mu\nu} h_{\alpha\beta}\dot z^{\mu}\dot
z^{\nu}\dot z^{\alpha}\dot z^{\beta} +
\cdot\cdot\cdot\right]\nonumber\\
&\equiv&  - M{\displaystyle\int} d \theta + {\displaystyle\int}
d^4 y h_{\mu\nu}(y)T_{\rm M
}^{\mu\nu}(y)+\frac{1}{2}{\displaystyle\int} d^4 y
h_{\mu\nu}(y)h_{\alpha\beta}(y)Q_{\rm M}^{\mu\nu\alpha\beta}(y)+
\cdot\cdot\cdot\,,
\end{eqnarray}
\noindent where
\begin{eqnarray}
T_{\rm M}^{\mu\nu}(y) = - \frac{M}{2}{\displaystyle\int} d \theta
\dot z^{\mu}\dot z^{\nu}\delta^4(y-z(\theta))\,, \nonumber
\end{eqnarray}
\begin{eqnarray}
Q_{\rm M}^{\mu\nu\alpha\beta}(y) = \frac{M}{4}{\displaystyle\int}
d \theta \dot z^{\mu}\dot z^{\nu}\dot z^{\alpha}\dot
z^{\beta}\delta^4(y-z(\theta))\,, \qquad \dot{z}^{\mu} \equiv
\frac{d z^{\mu}(\theta)}{d\theta}\,. \nonumber
\end{eqnarray}
\noindent

The $\Sigma$-functional entering the generating functional of
Green functions (\ref{genernew}) thus takes the form
\begin{eqnarray}\label{sigma1}
\Sigma = S_{\rm FP} &+& \int d^4x
~K^{\mu\nu}D_{\mu\nu}^{\alpha}C_{\alpha} - \int d^4
x~\frac{L^{\gamma}}{2}
f^{\alpha\beta}_{~~~\gamma}C_{\alpha}C_{\beta} \nonumber\\
&+& J\int d^4 x \frac{\delta S_{\rm m}}{\delta
h_{\mu\nu}}{D}_{\mu\nu}^{\alpha}C^{\alpha} + \EuScript{J}\int d^4
x \frac{\delta S_{\rm M}}{\delta
h_{\mu\nu}}{D}_{\mu\nu}^{\alpha}C^{\alpha}\,.
\end{eqnarray}
\noindent In this new formulation, the source $T^{\mu\nu}$
entering the generating functional is an auxiliary quantity set
equal to zero (together with all other sources) in the end of
calculations, and no longer plays the role of the source of the
gravitational field being measured.

To derive the Slavnov identities for the functional
$\mathfrak{W}\,,$ we perform the ``quantum'' BRST change of
integration variables in the functional integral (\ref{genernew}),
and obtain the following Slavnov identity
\begin{eqnarray}&&\label{slavalt}
{\displaystyle\int}dh dC d\bar{C} \left[\frac{\delta S_{\rm
m}}{\delta h_{\mu\nu}} D^{\alpha}_{\mu\nu} C_{\alpha} +
\frac{\delta S_{\rm M}}{\delta h_{\mu\nu}} D^{\alpha}_{\mu\nu}
C_{\alpha} + Y \bar{C}^{\alpha} F_{\alpha}^{,\mu\nu}
D^{\beta}_{\mu\nu} C_{\beta} + \frac{Y}{\xi} F_{\alpha}^{2} +
T^{\mu\nu}D^{\alpha}_{\mu\nu}C_{\alpha}  \right. \nonumber\\&&
\left. - \frac{\bar{\beta}^{\alpha}}{2}
f^{\alpha\beta}_{~~~\gamma}C_{\alpha}C_{\beta} -
\beta_{\alpha}\frac{F^{\alpha}}{\xi} \right] \exp\{i (\Sigma + Y
F_{\alpha}\bar{C}^{\alpha} + \bar{\beta}^{\alpha}C_{\alpha} +
\bar{C}^{\alpha}\beta_{\alpha} + T^{\mu\nu}h_{\mu\nu})\} = 0\,.
\end{eqnarray}
\noindent The first two terms in the square brackets will be
replaced by the derivatives with respect to the sources $J$ and
$\EuScript{J},$ respectively. The third term can be transformed
using the ghost equation of motion
\begin{eqnarray}\label{ghosteq}&&
{\displaystyle\int}dh dC d\bar{C}
\left[F_{\gamma}^{,\mu\nu}D_{\mu\nu}^{\alpha}C_{\alpha} - Y
F_{\gamma} + \beta_{\gamma} \right] \nonumber\\&& \times\exp\{i
(\Sigma + Y F_{\alpha}\bar{C}^{\alpha} +
\bar{\beta}^{\alpha}C_{\alpha} + \bar{C}^{\alpha}\beta_{\alpha} +
T^{\mu\nu}h_{\mu\nu})\} = 0\,, \nonumber
\end{eqnarray}
\noindent obtained by performing a shift $\bar{C} \to \bar{C} +
\delta\bar{C}$ of integration variables in the functional integral
(\ref{genernew}). Differentiating this equation with respect to
$\beta_{\gamma}$ gives
\begin{eqnarray}\label{ghosteq1}&&
Y {\displaystyle\int}dh dC d\bar{C} \left[i
\bar{C}^{\gamma}F_{\gamma}^{,\mu\nu}D_{\mu\nu}^{\alpha}C_{\alpha}
+ \beta_{\gamma}\frac{\delta}{\delta\beta_{\gamma}}\right]
\exp\{\cdot\cdot\cdot\} = 0\,, \nonumber
\end{eqnarray}
\noindent where the use of the property $Y^2 = 0$ has been made,
and the expression $\delta\beta_{\gamma}/\delta\beta_{\gamma} \sim
\delta^4(0),$ equal to zero in the dimensional regularization, is
omitted. Expressing the remaining terms as derivatives with
respect to the BRST-sources, and introducing the generating
functional of connected Green functions yields
\begin{eqnarray}\label{slavnovalt}&&
T^{\mu\nu}\frac{\delta \mathfrak{W}}{\delta K^{\mu\nu}} -
\bar{\beta}^{\alpha}\frac{\delta \mathfrak{W}}{\delta L^{\alpha}}
- \frac{1}{\xi} \beta_{\alpha}F^{\alpha,\mu\nu}\frac{\delta
\mathfrak{W}}{\delta T^{\mu\nu}} + \frac{\partial
\mathfrak{W}}{\partial J} + \frac{\partial \mathfrak{W}}{\partial
\EuScript{J}} - Y \beta_{\alpha}\frac{\delta
\mathfrak{W}}{\delta\beta_{\alpha}} - 2 Y\xi\frac{\partial
\mathfrak{W}}{\partial\xi} = 0\,.
\end{eqnarray}
\noindent Doing as in the preceding section, we obtain from
Eq.~(\ref{slavnovalt}) the following equation for the full
$\xi$-derivative of the effective device action:
\begin{eqnarray}\label{slavnov2alt}
2 \xi\frac{d\Gamma_{\rm m}}{d\xi} = -
\frac{\partial\EuScript{W}_{\rm m}}{\partial \EuScript{J}} -
\frac{\partial\EuScript{W}_{\rm m}}{\partial J} \,.
\end{eqnarray}
\noindent

Evaluation of the right hand side of Eq.~(\ref{slavnov2alt})
proceeds in the same way as in the case of a linear source, except
that there appear additional diagrams containing the $Q_{\rm
M}$-vertex. Namely, contribution of the second term is represented
by the diagrams of Fig.~\ref{fig2} with the substitution
$T^{\mu\nu} \to T^{\mu\nu}_{\rm M},$ and three new diagrams
pictured in Fig.~\ref{fig3}. As before, it is not difficult to
show that the sum of diagrams \ref{fig2}(c,d) and \ref{fig3}(c) is
equal to the diagram \ref{fig1}(d) in which the $K$-vertex is
substituted by the $\EuScript{J}$-vertex. Indeed, neglecting
device contribution, differentiating Eq.~(\ref{slavnovalt}) with
respect to $\beta_{\alpha},$ and setting all the sources equal to
zero gives
\begin{eqnarray}\label{slavnov5}
\frac{1}{\xi}F^{\alpha,\mu\nu} \frac{\delta W}{\delta T^{\mu\nu}}
= \frac{\delta}{\delta\beta_{\alpha}}\frac{\partial W}{\partial
\EuScript{J}}\,.
\end{eqnarray}
\noindent In the linear approximation with respect to the source
$M,$ the left hand side of this identity is represented by the sum
of the vertical parts of diagrams \ref{fig2}(c,d), \ref{fig3}(c),
while the right hand side -- by the horizontal part of
\ref{fig1}(d) (with $K\to \EuScript{J}$). It remains only to apply
Eq.~(\ref{slavnov4}) to the vertical part of the latter diagram.

The analytical expressions of the other two diagrams in
Fig.~\ref{fig3} have the form
$$\mathfrak{I}_j = \int\frac{d^4 p}{(2\pi)^4} I_j(p)\,,$$
where
\begin{eqnarray}
I_{3(a)}(p) &=&
     -i~\mu^{\varepsilon}
{\displaystyle\int}\frac{d^{4-\varepsilon}k}{(2\pi)^4} T_{\rm
m}^{\xi\eta}(p)p_{\xi} Q_{\rm M}^{\tau\rho\sigma\lambda}(-p)
    G_{\tau\rho\chi\theta}(k)
\tilde{G}_{\eta\delta}(p)
     \tilde{G}^{\alpha}_{\beta}(p+k)
     \xi
     \tilde{G}^{\beta\gamma}(p+k)\nonumber\\&&
     \times\{\eta_{\sigma\gamma}(p_{\lambda} + k_{\lambda})
     + \eta_{\lambda\gamma}(p_{\sigma} + k_{\sigma})\}
        \left\{
p^{\delta}\eta^{\mu\nu}-p^{\mu}\eta^{\delta\nu}-p^{\nu}\eta^{\delta\mu}
    \right\}
\nonumber\\&&
    \times
    \left\{
        k_{\alpha}\delta_{\mu\nu}^{\chi\theta} -
\delta_{\mu\alpha}^{\chi\theta}
        (p_{\nu}+k_{\nu})-\delta_{\nu\alpha}^{\chi\theta}(p_{\mu}+k_{\mu})
    \right\}\,,
\nonumber\\
I_{3(b)}(p) &=& i~\mu^{\varepsilon}
{\displaystyle\int}\frac{d^{4-\varepsilon}k}{(2\pi)^4}(p)Q_{\rm
M}^{\chi\theta\sigma\lambda}(-p)
    \tilde{G}^{\alpha}_{\beta}(p+k)
    \xi
    \tilde{G}^{\beta\gamma}(p+k)
    \nonumber\\&&
    \times G_{\chi\theta\omega\xi}(k)
    \{\eta_{\sigma\gamma}(p_{\lambda} + k_{\lambda})
     + \eta_{\lambda\gamma}(p_{\sigma} +
     k_{\sigma})\}\nonumber\\&&
     \times \left\{T_{\rm m}^{\mu\nu}(p)\left[
k_{\alpha}\delta_{\mu\nu}^{\omega\xi}
-\delta_{\mu\alpha}^{\omega\xi}(p_{\nu}+k_{\nu})
-\delta_{\nu\alpha}^{\omega\xi}(p_{\mu}+k_{\mu})
    \right] \right.\nonumber\\&&
    \left. - Q_{\rm m}^{\omega\xi\mu\nu}(p)
    \left[\eta_{\mu\alpha}(p_{\nu} + k_{\nu})
    + \eta_{\nu\alpha}(p_{\mu} + k_{\mu})\right]\right\}\,. \nonumber
\end{eqnarray}
\noindent As to the first term in the right hand side of
Eq.~(\ref{slavnov2alt}), its contribution is given by
Eqs.~(\ref{1a})--(\ref{1d}) with the substitution $T^{\mu\nu}\to
T^{\mu\nu}_{\rm M},$ plus the same expressions in which the factor
$T_{\rm M}^{\mu\nu}D^{\gamma,\alpha\beta}_{\mu\nu}$ is changed to
$Q_{\rm M}^{\mu\nu\alpha\beta}D^{(0)\gamma}_{\mu\nu}\,.$
Distinguishing these additional contributions by a prime, and
doing the loop integrals as in the preceding section we find
\begin{eqnarray}\label{form1}
I_j^{\rm log}(p) =
\frac{mM\xi}{16\pi^2}\ln\left(\frac{-p^2}{\mu^2}\right)\int
d\theta\int d\tau \exp\left\{ip_{\mu}[x^{\mu}(\tau) -
z^{\mu}(\theta)]\right\} \EuScript{I}_j(\tau,\theta,p)\,,
\end{eqnarray}
\noindent where
\begin{eqnarray}
\EuScript{I}^{\prime}_{1(a)}(\tau,\theta,p) &=&
\frac{1}{48}\left\{(2\xi - 5)(\dot{x}^{\mu}\dot{z}_{\mu})^2 +
2(2\xi+1)(\dot{x}^{\mu}\dot{z}_{\mu})^4\right\}\,,
\nonumber\\
\EuScript{I}^{\prime}_{1(b)}(\tau,\theta,p) &=&
\frac{1}{12}\left\{- 2
(\dot{x}^{\mu}\dot{z}_{\mu})^2 + 1\right\}\,, \nonumber\\
\EuScript{I}^{\prime}_{1(c)}(\tau,\theta,p) &=& -
\frac{1}{48}\left\{(2\xi - 3) +
2(4\xi-1)\frac{(\dot{x}^{\mu}p_{\mu})^2}{p^2} + 4(\xi +
1)(\dot{x}^{\mu}\dot{z}_{\mu})^2
\right\}\,, \nonumber\\
\EuScript{I}^{\prime}_{1(d)}(\tau,\theta,p) &=& 0\,, \nonumber
\end{eqnarray}
\begin{eqnarray}
\EuScript{I}_{3(a)}(\tau,\theta,p) &=& \frac{- 4\xi +
1}{24}\frac{(\dot{x}^{\mu}p_{\mu})^2}{p^2}\,, \nonumber\\
\EuScript{I}_{3(b)}(\tau,\theta,p) &=& \frac{2\xi +
1}{48}\left\{2(\dot{x}^{\mu}\dot{z}_{\mu})^4 -
(\dot{x}^{\mu}\dot{z}_{\mu})^2 - 1\right\}\,, \nonumber
\end{eqnarray}
\noindent and we have taken into account that
$\eta_{\mu\nu}\dot{z}^{\mu}\dot{z}^{\nu} = 1,$ $\ddot{z}^{\mu} =
0.$ Adding up all the contributions, we finally arrive at the
following expression for the $\xi$-derivative of the one-loop
correction to the effective action of a point testing particle in
the gravitational field of a point classical mass
\begin{eqnarray}&&
\frac{d\Gamma^{\rm loop}_{\rm m}}{d\xi} = \frac{mM}{32\pi^2}\int
d\tau \int d\theta\int \frac{d^4
p}{(2\pi)^4}\exp\{ip_{\mu}[x^{\mu}(\tau)- z^{\mu}(\theta)]\}
\ln\left(\frac{-p^2}{\mu^2}\right)\left\{\left(\frac{\xi}{6} -
\frac{1}{4}\right) \right.\nonumber\\&&\left. + \frac{5(2\xi -
1)}{24}(\dot{x}^{\mu}\dot{z}_{\mu})^2 + \frac{5(2\xi -
1)}{6}\frac{(\dot{x}^{\mu}p_{\mu})^2}{p^2} - \frac{1}{6}
\frac{(\dot{x}^{\alpha}p_{\alpha})^2}{p^2}
(\dot{x}^{\mu}\dot{z}_{\mu})^2 + \frac{2\xi + 1
}{12}(\dot{x}^{\mu}\dot{z}_{\mu})^4\right\}\,. \nonumber
\end{eqnarray}\noindent
In the case when the particle producing gravitational field is at
rest, and the testing particle is non-relativistic, one has
$z^{\mu}(\theta) = (\theta, \bm{z}_0),$
$\dot{x}^{\mu}\dot{z}_{\mu} \approx 1,$ and an elementary
integration gives
\begin{eqnarray}&&\label{result2}
\frac{d\Gamma^{\rm loop}_{\rm m}}{d\xi} = mM\frac{3(2\xi - 1)
}{256\pi^2}\int d t \int \frac{d^3
\bm{p}}{(2\pi)^3}\exp\{-i\bm{p}[\bm{x}(t)- \bm{z}_0]\}
\ln\left(\frac{\bm{p}^2}{\mu^2}\right)\nonumber\\&& = - mM\frac{3
(2\xi - 1)}{512\pi^3}\int \frac{dt}{|\bm{x}(t)- \bm{z}_0|^3}\,.
\end{eqnarray}
\noindent As in the case of a linear source, the value of the
static gravitational potential turns out to be gauge-dependent. In
the ordinary units,
\begin{eqnarray}&&\label{main}
\frac{d V}{d\xi} = \frac{3(2\xi - 1)}{2\pi}\frac{G^2\hbar M}{c^3
r^3}\,, \qquad r = |\bm{x} - \bm{z}_0|\,.
\end{eqnarray}
\noindent This completes exposition of the main results of the
work.

\section{Discussion and conclusions}\label{conclud}

\begin{figure}
\scalebox{0.5}{\includegraphics{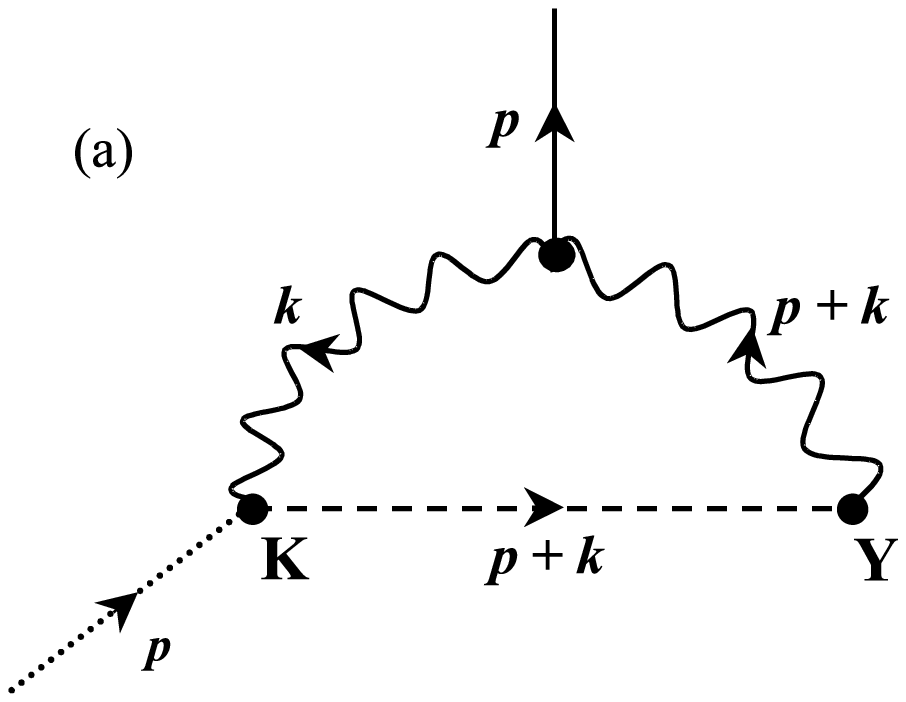}}
\scalebox{0.5}{\includegraphics{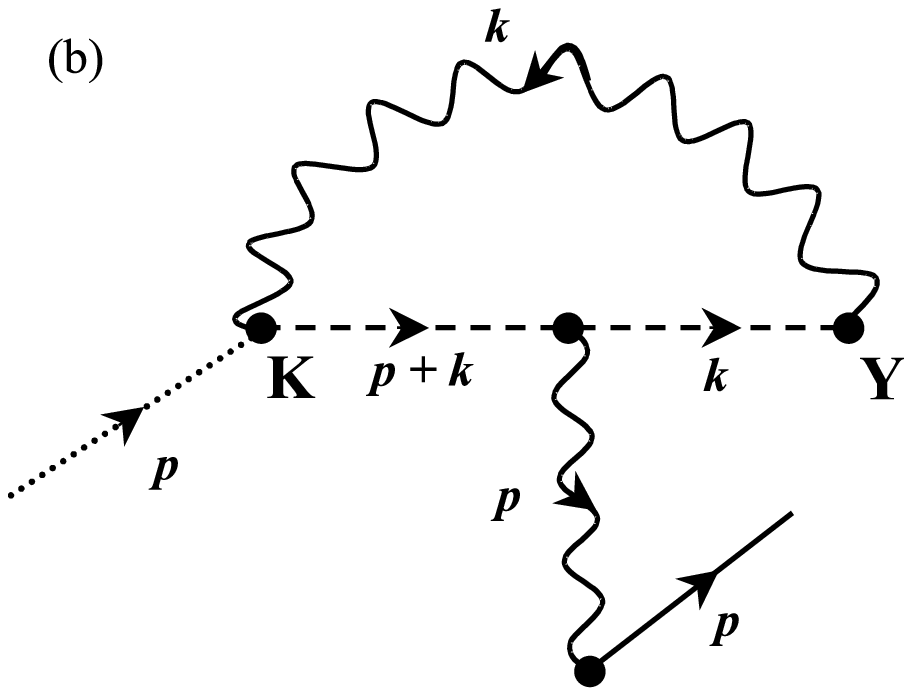}}
\scalebox{0.5}{\includegraphics{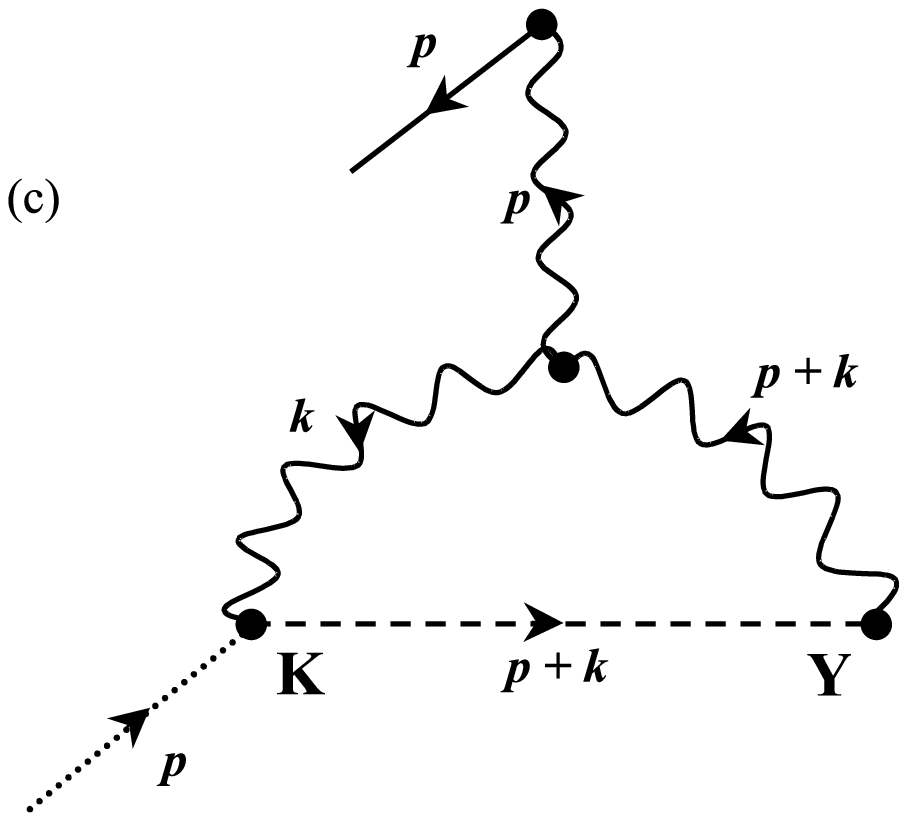}}
\scalebox{0.5}{\includegraphics{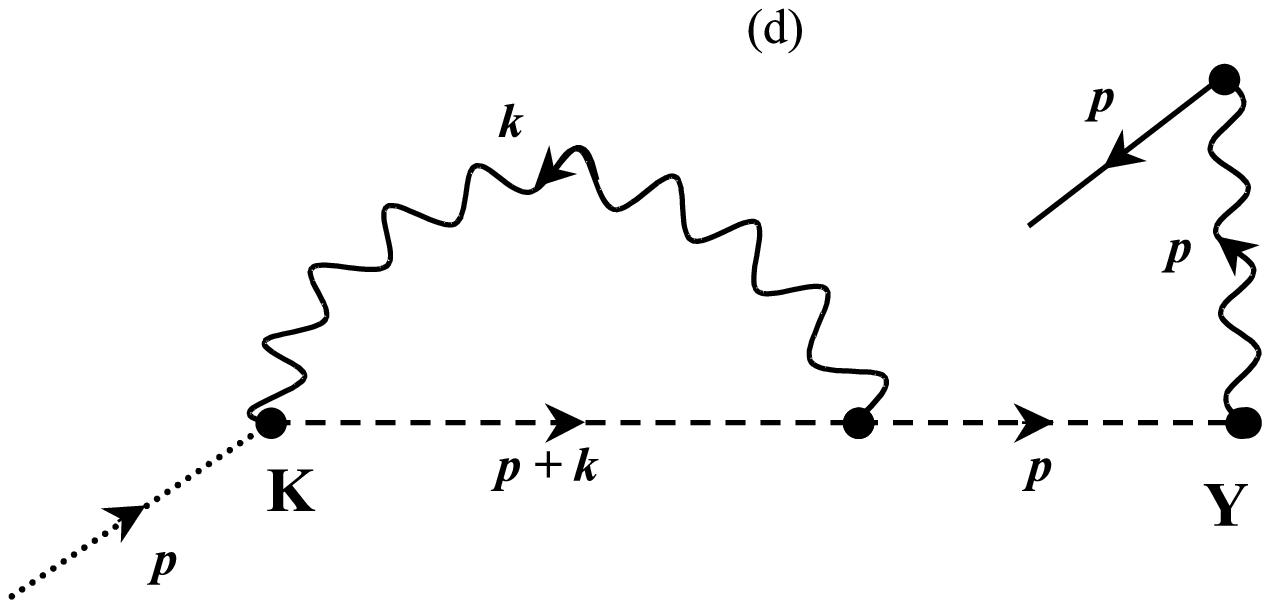}}
\scalebox{0.5}{\includegraphics{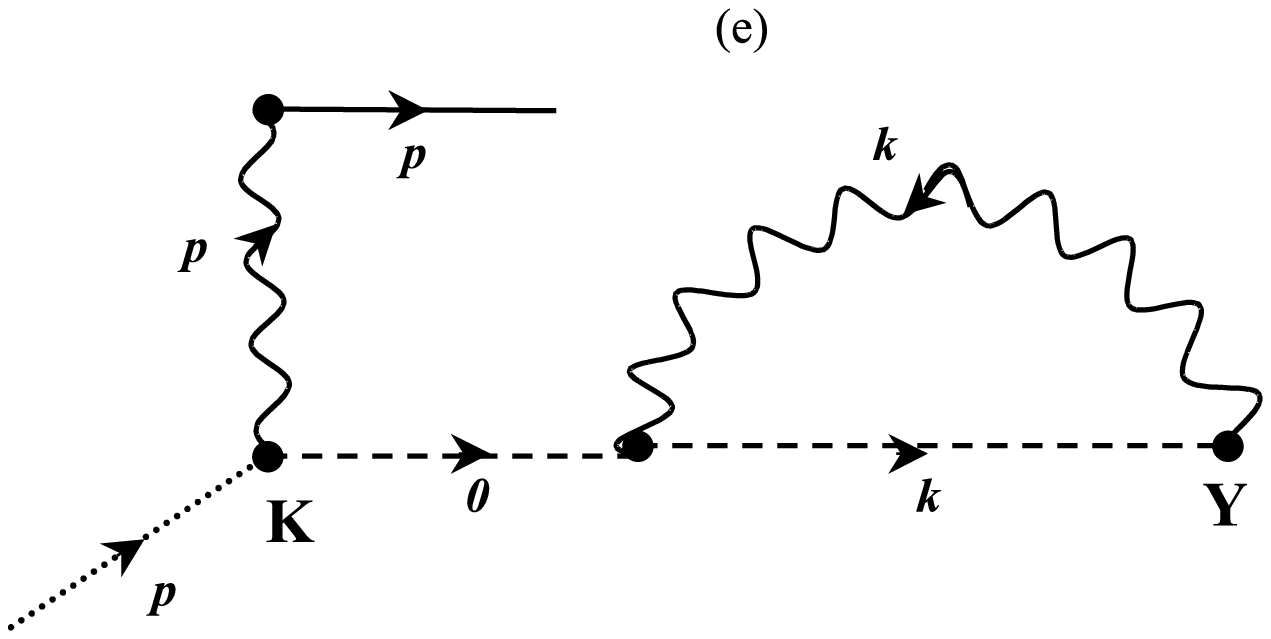}}
\scalebox{0.4}{\includegraphics{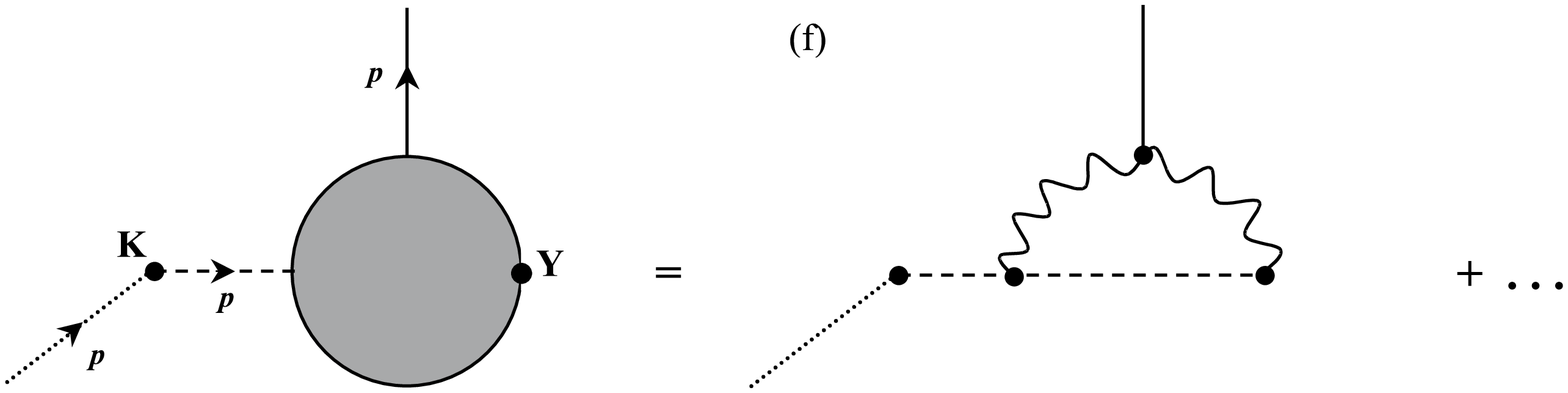}} \caption{The one-loop
contribution of the first term in the right hand side of
Eq.~(\ref{slavnov2}). Wavy lines represent gravitons, broken lines
ghosts, dotted lines the source $T^{\mu\nu},$ solid lines denote
the coefficient functions $T_{\rm m}^{\mu\nu},$ $Q_{\rm m
}^{\mu\nu\alpha\beta}$ appearing in the expansion of the device
action. Diagrams of the type (f) do not contribute in view of the
conservation law (\ref{conserv}), so we do not picture them in
detail.} \label{fig1}
\end{figure}

\begin{figure}
\scalebox{0.5}{\includegraphics{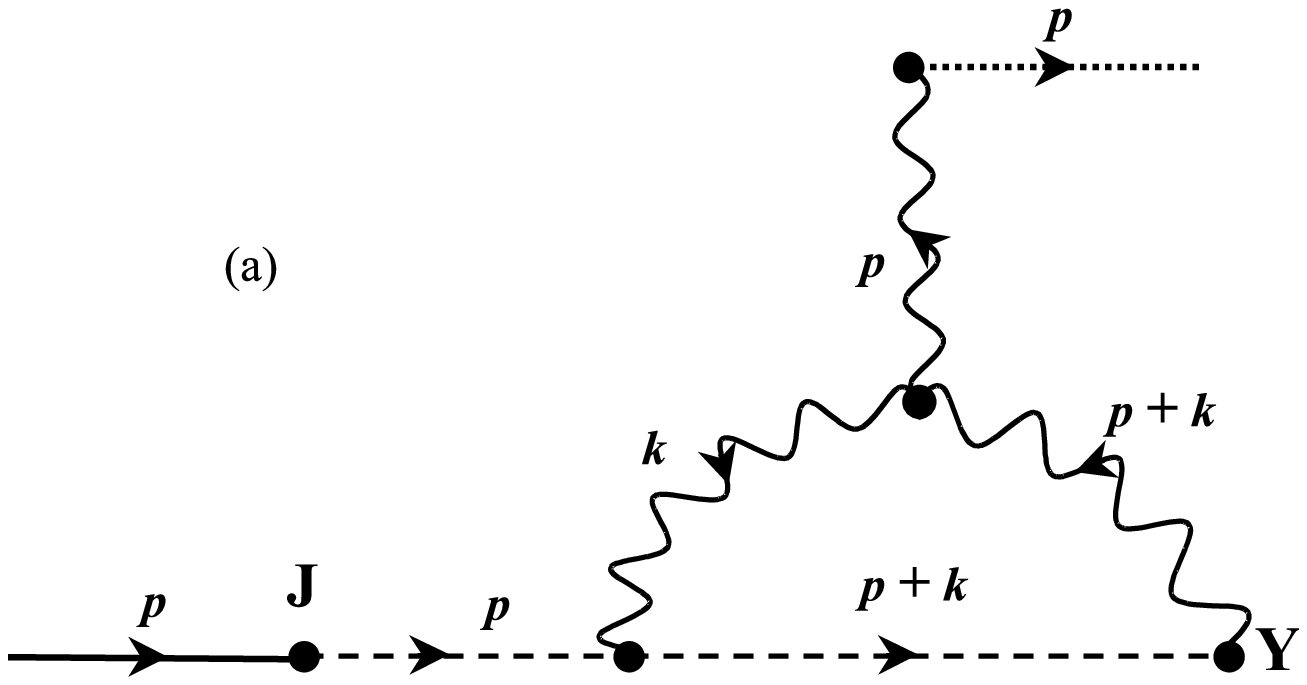}}
\scalebox{0.5}{\includegraphics{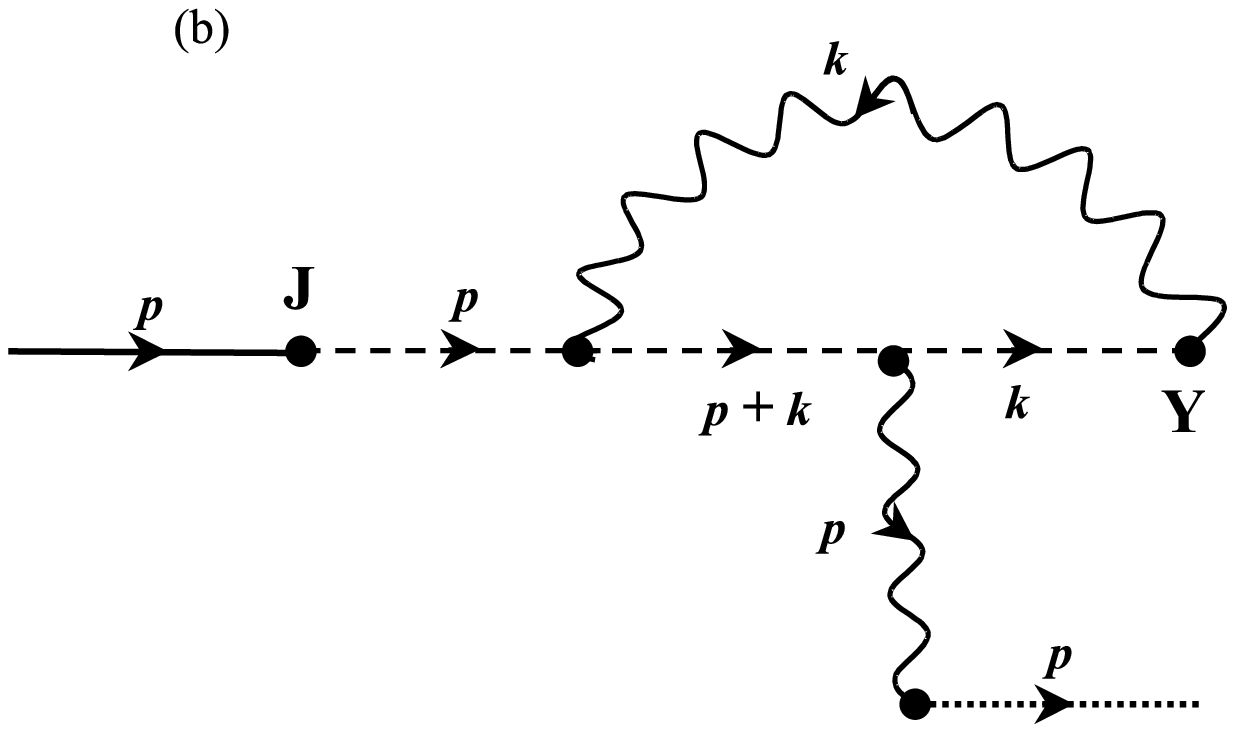}}
\scalebox{0.5}{\includegraphics{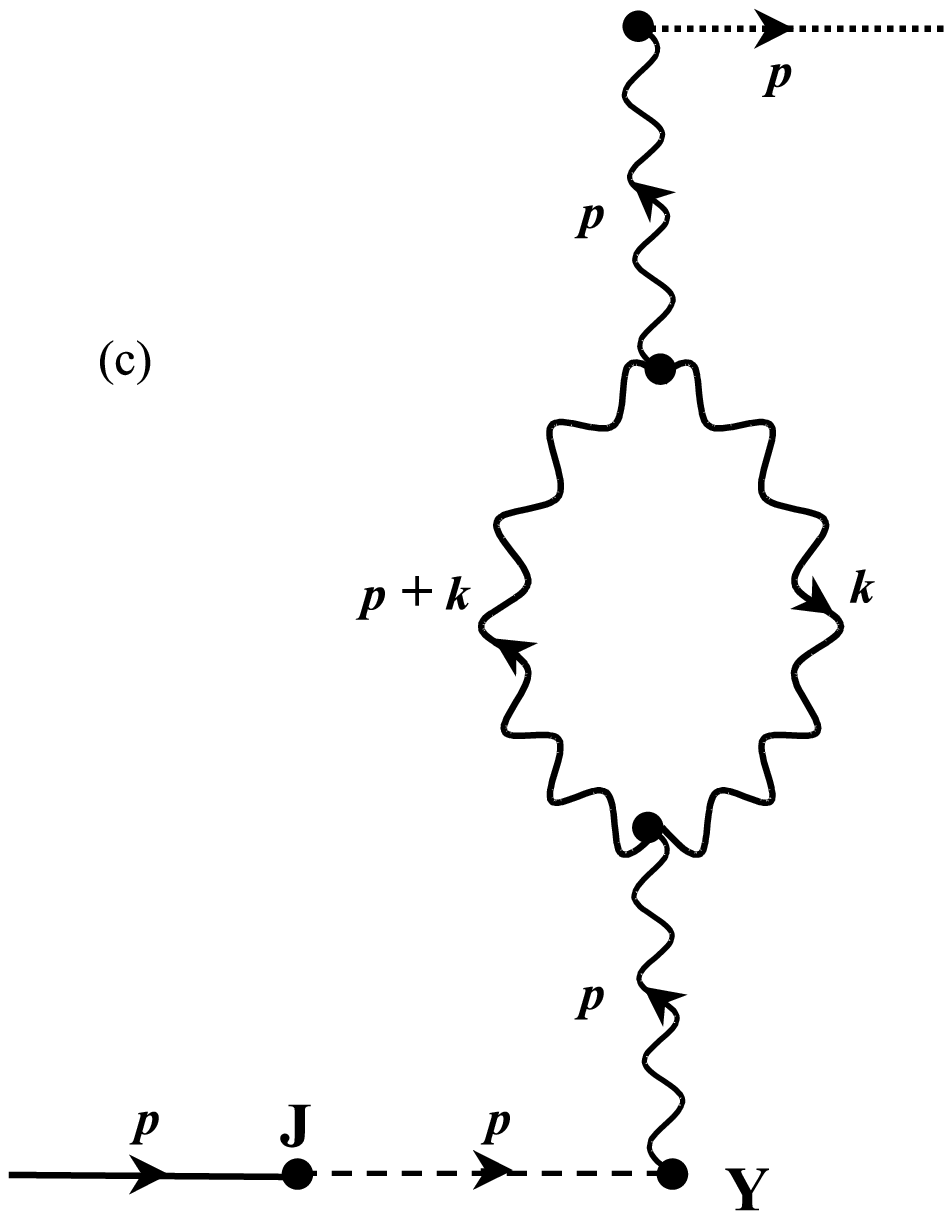}}
\scalebox{0.5}{\includegraphics{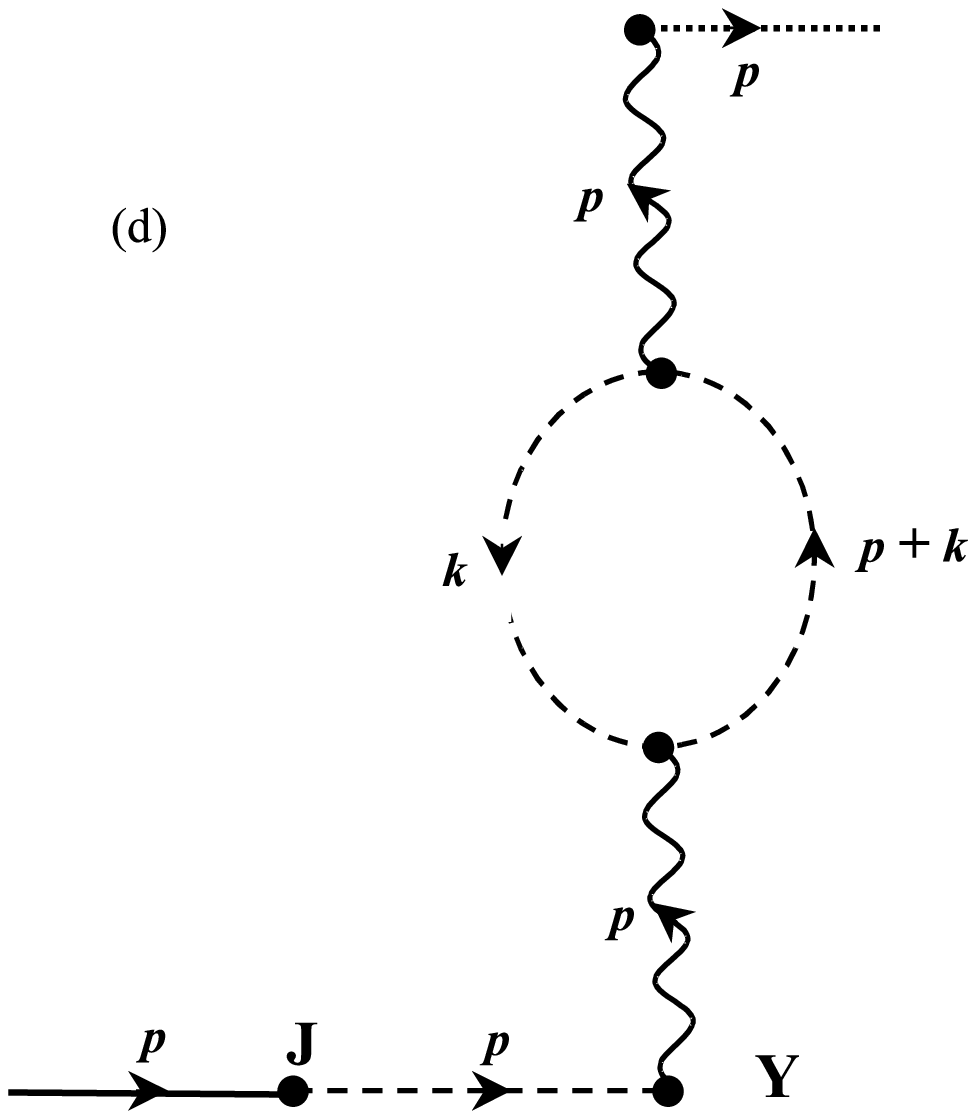}}
\scalebox{0.5}{\includegraphics{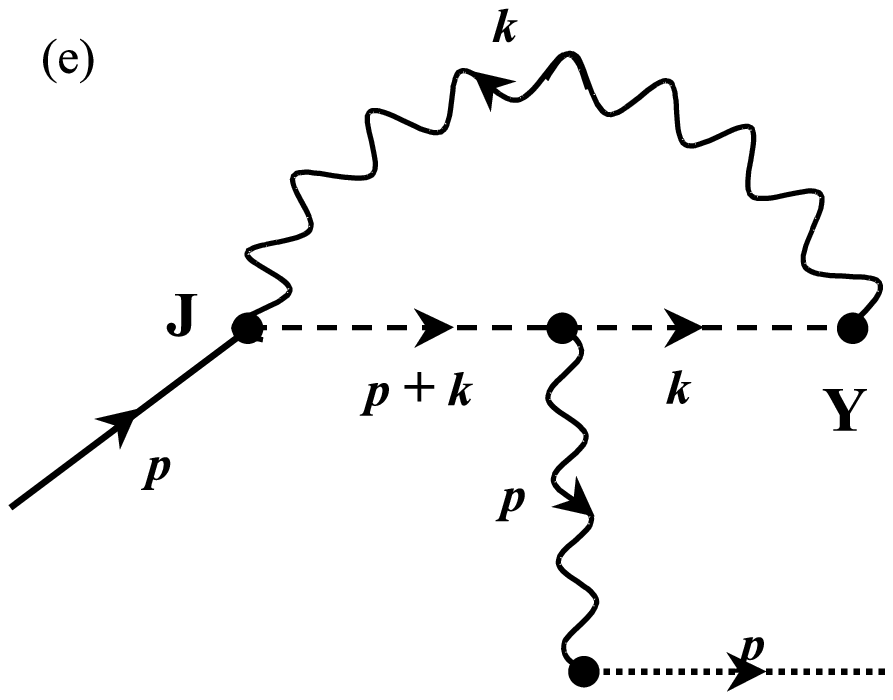}}
\scalebox{0.5}{\includegraphics{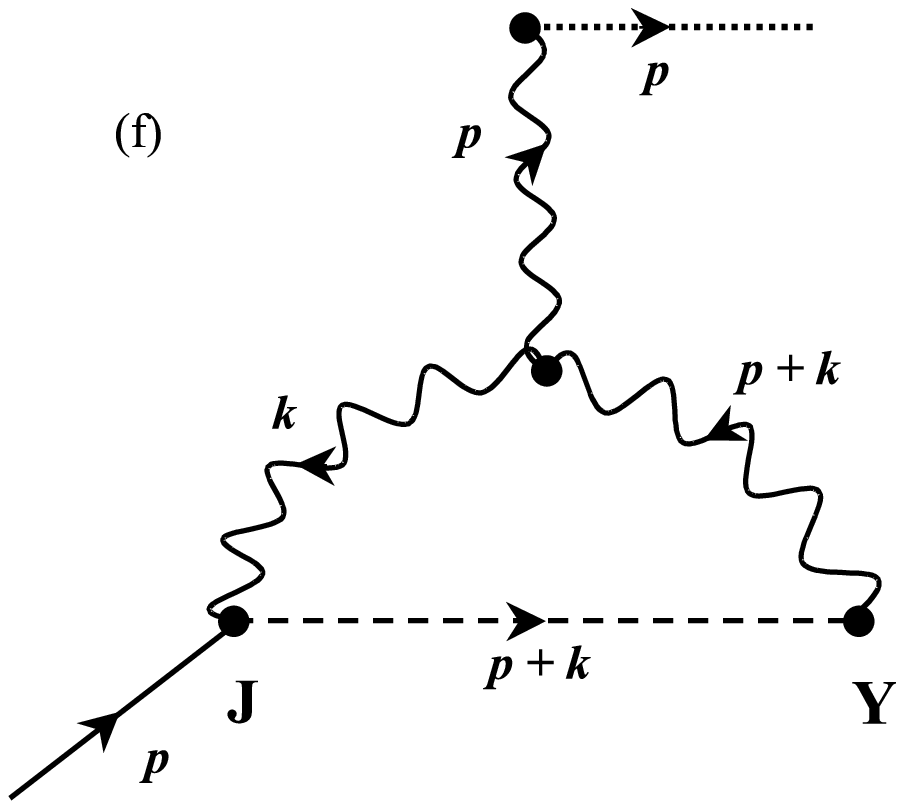}}
\scalebox{0.5}{\includegraphics{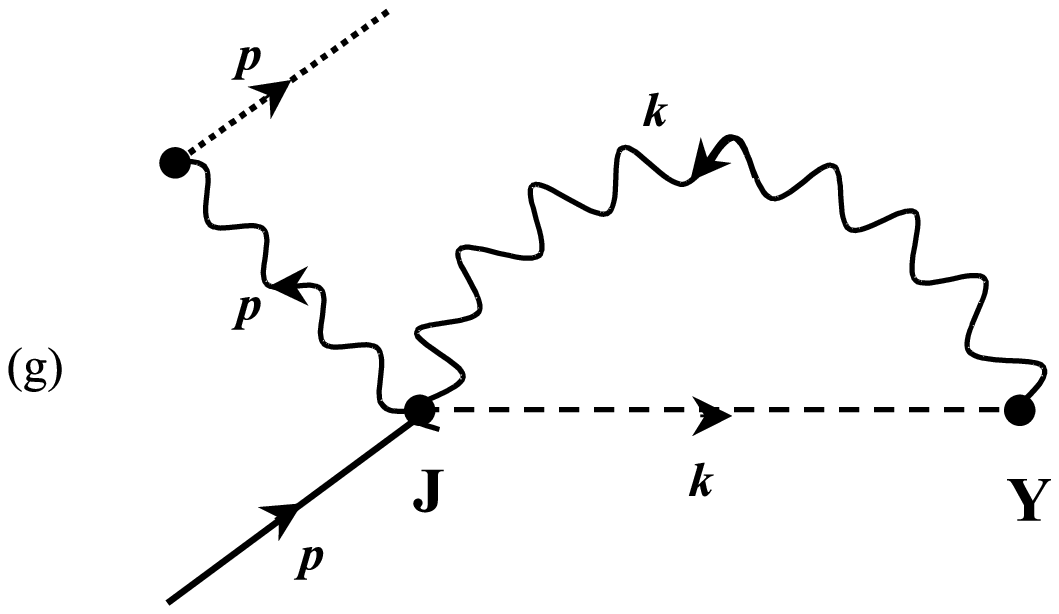}}
\scalebox{0.4}{\includegraphics{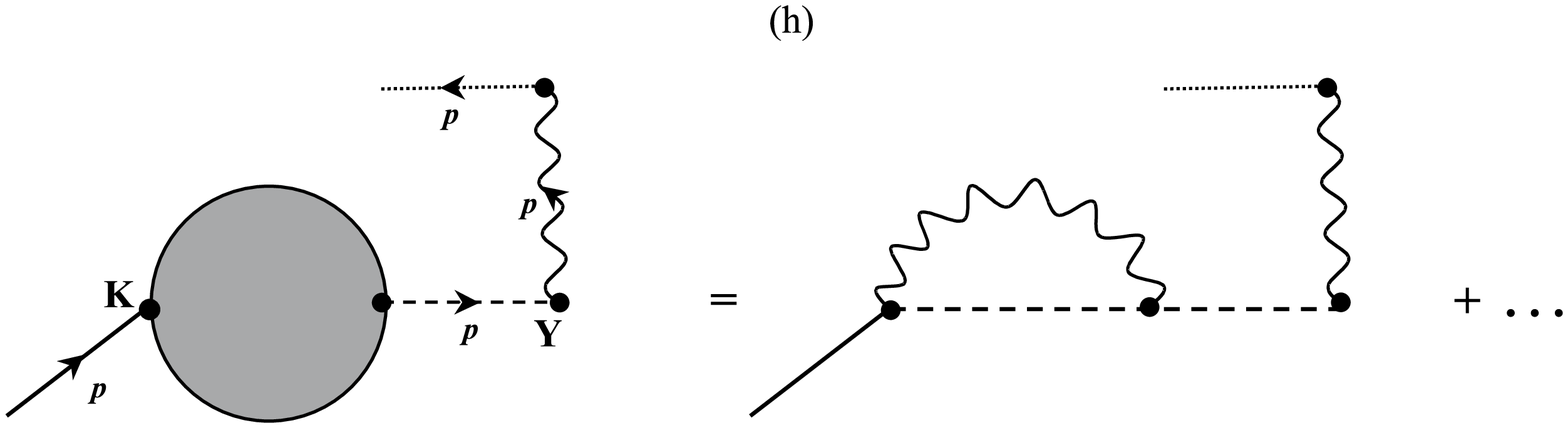}} \caption{The one-loop
contribution of the second term in the right hand side of
Eq.~(\ref{slavnov2}). Diagrams of the type (h) do not contribute
because of the energy-momentum conservation (\ref{conserv}).}
\label{fig2}
\end{figure}

\begin{figure}
\scalebox{0.5}{\includegraphics{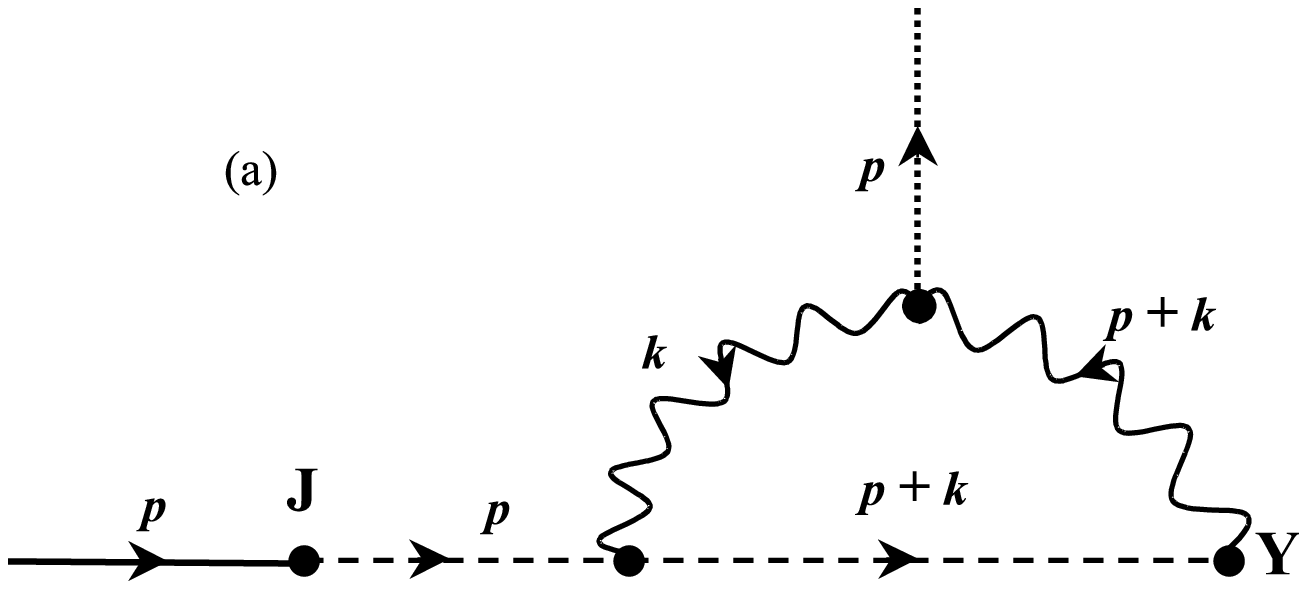}}
\scalebox{0.5}{\includegraphics{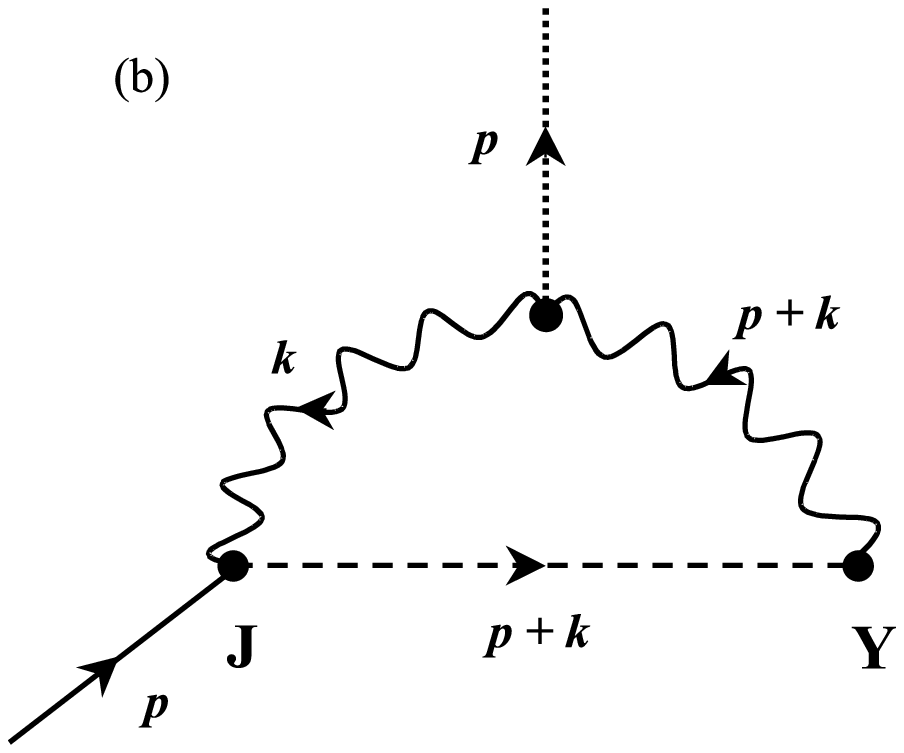}}
\scalebox{0.4}{\includegraphics{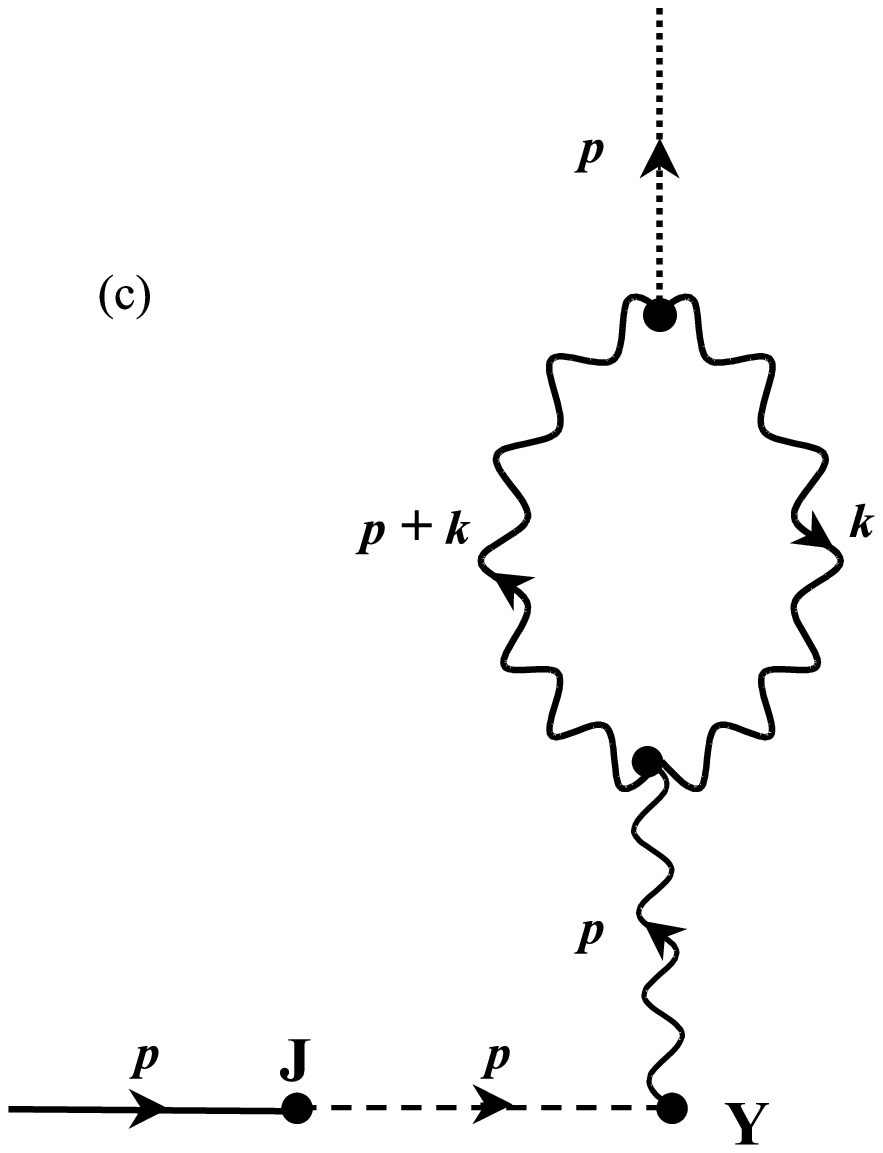}} \caption{Diagrams
accounting the nonlinearity of the graviton-source interaction,
generated by the second term in the right hand side of
Eq.~(\ref{slavnov2}).} \label{fig3}
\end{figure}

We have shown that, as in the case of scalar field considered in
I, the $O(\hbar)$ contribution to the effective action of a
point-like particle is gauge dependent, and that this dependence
leads to an ambiguity in the values of observables. This was
demonstrated in the case of an arbitrary linear source in
Sec.~\ref{linsource}, as well as in the particular case of a
nonlinear point-like source in Sec.~\ref{pointsource}. In our
opinion, these results are sufficient to conclude that the
explicit introduction of the classical measuring apparatus into
consideration does not solve the gauge dependence
problem.\footnote{This conclusion, of course, does not abolish the
fact that such an introduction is necessary as far as
contributions of the order $O(\hbar)$ are considered.} Still, some
remarks concerning the whole approach are worth to be made. The
method used in this paper allows one to elucidate the relative
role of the two factors, explicit and implicit, contributing to
the gauge dependent part of the effective device action. Let us
consider equation (\ref{slavnov2}) [or Eq.~(\ref{slavnov2alt}) in
the case of a nonlinear source]. Clearly, the first term in the
right hand side of this equation describes gauge dependence of the
mean gravitational field (taking into account contribution of the
nonlinear graviton-device interaction), {\it i.e.}, it represents
an implicit contribution to $\partial\Gamma_{\rm m}/\partial\xi,$
while the second term corresponds to an explicit gauge dependence
of $\Gamma_{\rm m}.$ The general analysis performed in
Secs.~\ref{linsource}, \ref{pointsource} reveals the fact that the
two contributions tend to add up, rather than subtract,
undermining thereby the very idea of the gauge dependence
cancellation. Indeed, as we have shown using the Slavnov
identities, contribution of the diagrams \ref{fig2}(c,d) [and
\ref{fig3}(c) in the nonlinear case] doubles the contribution of
diagram \ref{fig1}(d), rather than cancels it. However,
correlation between the rest of diagrams in Figs.~\ref{fig1},
\ref{fig2} is more intricate. Namely, it is not difficult to
verify that the sum of these diagrams is zero identically provided
that the diagrams \ref{fig1}(b) and \ref{fig2}(f) are taken with
opposite sings. Analogously, additional contributions found in
Sec.~\ref{pointsource} cancel each other if the signs of
$\EuScript{I}^{\prime}_{1(b)}$ and $\EuScript{I}_{3(a,b)}$ are
reversed. At the same time, it should be noted that even if the
two terms entered the right hand side of Eq.~(\ref{slavnov2}) with
opposite signs, the action $\Gamma_{\rm m}$ would still be
$\xi$-dependent.\footnote{In the case of a point-like source,
considered in Sec.~\ref{pointsource}, $\Gamma_{\rm m}$ would be
$\xi$-independent in the non-relativistic limit if the signs of
the two terms in Eq.~(\ref{slavnov2alt}) were opposite, but this
is a trivial consequence of the symmetry of these terms with
respect to the change $m\leftrightarrow M$ in this limit.}

Finally, we would like to note that although the main conclusion
of the present paper is the same as in I in the case of scalar
field, specific form of the $\xi$-dependence of the effective
gravitational field is different. Therefore, one should not
exclude the possibility that the resolution of the gauge
dependence problem can still be found in the context discussed
above by appropriately refining the model of the measurement
process.

{}

\pagebreak

\pagebreak

\pagebreak

\end{document}